\documentclass[fleqn,usenatbib]{mnras}

\usepackage{newtxtext,newtxmath}
\usepackage[T1]{fontenc}
\usepackage{ae,aecompl}

\usepackage{amsmath}
\usepackage{graphicx}
\usepackage{multirow}
\usepackage{multicol}
\usepackage{gensymb}
\usepackage{caption}
\usepackage{url}
\usepackage[utf8]{inputenc}
\usepackage[titletoc]{appendix}
\usepackage{array}% http://ctan.org/pkg/array
\usepackage{balance}
\usepackage{float}

\newcommand{\code}[1]{\texttt{#1}}

\title[The Nature of 500 Micron Risers II]{The Nature of 500 Micron Risers II: Sub-mm Faint Dusty Star-Forming Galaxies}

\author[J. Cairns et al.]{
J. Cairns$^{1}$\thanks{E-mail: j.cairns18@imperial.ac.uk},
D. L. Clements$^{1}$,
J. Greenslade$^{1}$,
G. Petitpas$^{2}$,
T. Cheng$^{1}$,
Y. Ding$^{1}$,
\newauthor{
A. Parmar$^{1}$,
I. P\'erez-Fournon$^{3,4}$
and D. Riechers$^{5}$}
\\
$^{1}$Imperial College London, Blackett Laboratory, Prince Consort Road, London, SW7 2AZ, UK\\
$^{2}$Harvard-Smithsonian Center for Astrophysics, 60 Garden Street, Cambridge, MA 02138\\
$^{3}$Instituto de Astrofísica de Canarias, C/Vía Láctea, s/n, E-38205 San Cristóbal de La Laguna, Tenerife, Spain\\
$^{4}$Universidad de La Laguna, Dpto. Astrofísica, E-38206 San Cristóbal de La Laguna, Tenerife, Spain\\
$^{5}$I. Physikalisches Institut, Universit\"at zu K\"oln, Z\"ulpicher Strasse 77, D-50937 K\"oln, Germany
}

\date{Accepted XXX. Received YYY; in original form ZZZ}

\pubyear{2021}

\begin{document}
\label{firstpage}
\pagerange{\pageref{firstpage}--\pageref{lastpage}}
\maketitle

\begin{abstract}
We present SCUBA-2 and SMA follow-up observations of four candidate high redshift Dusty Star-Forming Galaxies, selected as sources with rising SEDs in the 250, 350 and 500\,$\mu$m \textit{Herschel} SPIRE bands. Previous SMA observations showed no counterparts to these sources, but in our deeper sub-mm observations we detect counterparts to all four 500\,$\mu$m risers, with three resolving into multiple systems. For these three multiple systems, the SMA 345\,GHz ($\approx 870$\,$\mu$m) observations recover $123 \pm 73\%$, $60 \pm 15\%$ and $19 \pm 4\%$ respectively of the integrated 850\,$\mu$m flux density from SCUBA-2, indicating that there may be additional sources below our SMA detection limit making up a dense, protocluster core. The fourth 500\,$\mu$m riser was observed at a lower frequency and so we cannot make a similar comparison. We estimate photometric redshifts based on FIR/sub-mm colours, finding that $3/4$ likely lie at $z \geq 2$. This fits with the interpretation that the 500\,$\mu$m riser selection criterion selects both intrinsically red, individual galaxies at $z > 4$, and multiple systems at more moderate redshifts, artificially reddened by the effects of blending. We use the SCUBA-2 850\,$\mu$m maps to investigate the environments of these 500\,$\mu$m risers. By constructing cumulative number counts and estimating photometric redshifts for surrounding SCUBA-2 detections, we find that one of our 500\,$\mu$m risers could plausibly reside in a $z \geq 2$ protocluster. We infer that bright 500\,$\mu$m risers with faint 850\,$\mu$m flux densities are typically multiple systems at $z \geq 2$ that may reside in overdensities of bright sub-mm galaxies.

\end{abstract}

\begin{keywords}
galaxies: starburst; galaxies: high-redshift; submillimeter: galaxies; infrared: galaxies
\end{keywords}

%%%%%%%%%%%%%%%%%%%%%%%%%%%%%%%%%%%%%%%%%%%%%%%%%%

%%%%%%%%%%%%%%%%% BODY OF PAPER %%%%%%%%%%%%%%%%%%

\section{Introduction}
\label{sec: intro}

Dusty Star-Forming Galaxies (DSFGs) provide an enormous contribution to the total energy density emitted by galaxies and hence represent a vital stage in their formation and evolution. In the last couple of decades, there has been a profound development in our understanding of the nature of DSFGs and their role in galaxy formation and evolution, particularly those selected at sub-mm wavelengths (typically 850\,$\mu$m). These sources have subsequently come to be known as sub-mm galaxies or SMGs \citep{2002PhR...369..111B}. The redshift distribution of SMGs peaks at $z \sim 2 - 3$ with a significant tail extending to higher redshifts \citep{2005ApJ...622..772C,2011MNRAS.415.1479W,2014ApJ...788..125S,2014MNRAS.444..117K,2017MNRAS.471.2453S,2017ApJ...840...78D,2019MNRAS.487.4648S}. SMGs are typically massive \cite[$M_{\star} \sim 10^{10} - 10^{11}$\,M$_{\odot}$:][]{2004ApJ...617...64S,2011ApJ...740...96H,2012A&A...541A..85M,2013MNRAS.432...23G,2014ApJ...788..125S,2020MNRAS.494.3828D,2021MNRAS.504..928P}), rich in both gas and dust \citep{2002MNRAS.331..495S,2005MNRAS.359.1165G,2006ApJ...650..592K,2006ApJ...640..228T,2008ApJ...680..246T,2010ApJ...714.1407C,2010MNRAS.403..274C,2013MNRAS.429.3047B,2020MNRAS.494.3828D,2021MNRAS.504..928P} and extremely luminous \cite[with total infrared luminosities $L_{\text{TIR}} \sim 10^{11} - 10^{13}$\,L$_{\odot}$:][]{2005ApJ...622..772C,2012A&A...548A..22M,2014MNRAS.438.1267S,2015MNRAS.451.3419G,2017MNRAS.468.4006M,2017MNRAS.469..492M,2018A&A...619A.169R,2018MNRAS.477.2042H,2019MNRAS.490.3840C,2020MNRAS.496.2315G}, corresponding to dust-obscured star formation rates (SFRs) of hundreds to thousands of solar masses per year and hence contributing significantly to the Cosmic Star Formation Rate Density (CSFRD) at these redshifts \citep{2005ApJ...622..772C,2011MNRAS.415.1479W,2012ApJ...761...89B,2014MNRAS.438.1267S,2017MNRAS.471.2453S,2020MNRAS.494.3828D}. There is still some debate surrounding the morphology and origin of DSFGs and they are likely a diverse population in this respect \cite[see e.g.][]{2014PhR...541...45C}. Some studies suggest that many DSFGs may represent isolated, gas-rich, disk galaxies undergoing a burst of star formation \citep{2010ApJ...714.1407C,2010MNRAS.404.1355D,2013MNRAS.432.2012T,2017MNRAS.469..492M}, while other studies suggest that they are primarily merger-driven \citep{10.1111/j.1365-2966.2004.08553.x,2008ApJ...680..246T,2010ApJ...724..233E,2012MNRAS.425.1320I,2012ApJ...757...23K,2015ApJ...799..194C,2021MNRAS.503.2622C} and potentially represent a key, but short-lived  \cite[typically of order a few hundred Myr:][]{2016ApJ...824...36C,2021MNRAS.504..928P} phase in an evolutionary sequence, followed by a bright quasar phase which then decays to leave a massive elliptical galaxy \citep{2005ApJ...632..736A,2006ApJ...641L..17F,2008ApJS..175..356H,2010MNRAS.402.2113C,2014ApJ...782...68T,2017MNRAS.464.1380W,2020MNRAS.494.3828D}. Whilst SMGs are important objects to study in their own right, a number of studies have found numerous DSFGs residing within high redshift protoclusters \citep{2009ApJ...694.1517D, 2011Natur.470..233C, 2012Natur.486..233W, 2014MNRAS.439.1193C, 2014A&A...570A..55D, 2015ApJ...815L...8U, 2015ApJ...808L..33C, 2016MNRAS.461.1719C, 2016ApJ...828...56W, 2018MNRAS.476.3336G, 2018ApJ...856...72O, 2018Natur.556..469M, 2019A&A...625A..96K, 2019ApJ...872..117G, 2020arXiv201002909W, 2021A&A...646A.174A, 2021ApJ...916...46J} thought to be the progenitors of local, massive galaxy clusters. SMGs therefore additionally have the potential to be signposts for these overdense regions in the early Universe, whose study can be used to test cosmological models and investigate the formation and evolution of large scale structure \cite[see e.g.][]{2018Natur.553...51M}. 

Despite this immense progress, our understanding of the nature of SMGs remains incomplete, particularly for those objects residing at $z > 4$ where small samples and a lack of spectroscopic confirmations obstructs further progress. The \textit{Herschel Space Observatory} discovered a surprisingly large population ($3.3 \pm 0.8$\,deg$^{-2}$) of extremely luminous, candidate $z > 4$ DSFGs, selected as sources with a rising Spectral Energy Distribution (SED) in the three \textit{Herschel} SPIRE \citep{2010A&A...518L...3G} bands \cite[i.e. $S_{250} < S_{350} < S_{500}$:][]{2014ApJ...780...75D, 2016MNRAS.462.1989A,2016ApJ...832...78I} and hence typically known as `500\,$\mu$m risers'. Numerous 500\,$\mu$m risers have been confirmed at $z > 4$ \cite[e.g.][]{2011ApJ...740...63C, 2013Natur.496..329R, 2017ApJ...850....1R, 2021ApJ...907...62R, 2017MNRAS.472.2028F} with corresponding SFRs in excess of 1000\,M$_{\odot}$yr$^{-1}$, making them extreme objects, potentially at the high luminosity end of a larger population of high-$z$ DSFGs \citep{2020MNRAS.496.2315G}. Simulations have significant difficulty in reproducing this population whilst simultaneously satisfying other observational constraints \cite[see e.g.][and references therein]{2021MNRAS.502.2922H}. Accumulating observations of high redshift DSFGs will therefore be vital in gaining a complete understanding of this population. 
Very few studies have focused on investigating the nature of these 500\,$\mu$m risers, and as such the majority of their properties remain poorly constrained. Simulations suggest that $\sim 40\%$ of faint (S$_{500} < 60$\,mJy) \textit{Herschel} sources should be comprised of multiple DSFGs, while the brightest should be exclusively strongly lensed single galaxies \citep{2017A&A...607A..89B}. Additionally, some fraction of these multiple systems will be chance line-of-sight alignments of DSFGs, rather than physically associated structures. \cite{2017arXiv170904191O} provide follow-up observations of a sample of 44 500\,$\mu$m risers with the Atacama Large Millimeter Array \cite[ALMA:][]{2004AdSpR..34..555B}, finding that $\sim 61\%$ resolve into a single sub-mm bright source, while the remainder break up into multiple DSFGs (four break up into $\geq 3$ sources, with one resolving into five separate sources in the synthesised $\sim 0.12^{\prime\prime}$ ALMA beam). Additionally, 18 of their sources show some evidence of gravitational lensing. \cite{2020MNRAS.496.2315G}, hereafter known as G20, selected a sample of thirty-four 500\,$\mu$m risers from the \textit{Herschel} Multi-tiered Extragalactic Survey \cite[HerMES:][]{2012MNRAS.424.1614O}, and carried out interferometric follow-up observations with the Submillimeter Array \cite[SMA:][]{2004ApJ...616L...1H} between 2010 and 2015. They found that 4 break up into two individual sources and 18 resolve into a single source, with the remaining 12 maps containing no apparent SMA counterpart. G20 suggested that non-detections were likely a result of the bright \textit{Herschel} sources breaking up into multiple faint counterparts clustered within the \textit{Herschel} SPIRE beamsize ($\sim 17.6^{\prime\prime}$ for the 250\,$\mu$m band) but separated by distances greater than the SMA beamsize ($\sim2^{\prime\prime}$) and lying below their detection threshold, with simple flux calculations indicating that these multiple systems should contain $\geq 3$ individual DSFGs. Based on this assumption, G20 estimated that $\sim 60\%$ of faint (S$_{500} < 60$\,mJy) 500\,$\mu$m risers and $\sim 35\%$ of bright (S$_{500} > 60$\,mJy) 500\,$\mu$m risers should be blends, indicating that this population is likely much more diverse than predicted. Similarly, based on LMT/AzTEC 1.1\,mm follow-up observations of their sample of bright \textit{Herschel} 500\,$\mu$m risers, \cite{2021MNRAS.505.5260M} find that $45/93$ have no counterpart in the higher resolution imaging, with some of these non detections expected to be multiple systems and others expected to be individual, resolved sources that are faint at 1.1\,mm due to a large dust spectral emissivity index $\beta$. The multiplicity of their sample could therefore be $\sim 9\%$ in the most conservative scenario, or $\sim 50-60\%$ in the most extreme scenario. Any systems with high multiplicities would naturally be candidate high redshift protocluster cores containing numerous DSFGs, structures discovered in only a handful of studies to date \cite[e.g.][]{2018Natur.556..469M, 2018ApJ...856...72O}.

In order to determine a more robust multiplicity fraction for samples of 500\,$\mu$m risers, deeper follow-up observations of 500\,$\mu$m risers without apparent cross-matches in higher resolution data are required. We therefore followed on from G20 and targeted four of their 500\,$\mu$m risers without an apparent SMA counterpart for additional, deeper FIR/sub-mm observations. Given the significant improvements to the SMA since 2015 (particularly in terms of the improved bandwidth provided by the SWARM correlator), we were able to obtain much deeper, high-resolution SMA 345\,GHz continuum imaging of these 500\,$\mu$m risers. We additionally obtained complementary observations at 850\,$\mu$m with the Submillimeter Common User Bolometer Array 2 \cite[SCUBA-2:][]{2013MNRAS.430.2513H} on the James Clerk Maxwell Telescope (JCMT), which provide an integrated flux density with a coarser resolution at a similar wavelength. The primary aim of this paper is to determine the multiplicities of these sources, allowing us to comment further on the diversity of the 500\,$\mu$m riser population. The SCUBA-2 observations also allow us to investigate the wider field surrounding these sources and hence evaluate the environments in which they reside.

This paper is structured as follows: In Section \ref{sec: sample} we describe the selection of the 500\,$\mu$m riser sample, followed by a discussion of the observations and data reduction in Section \ref{sec: bootes_observations}. In Sections \ref{sec: results} and \ref{sec: discussion} we present the results of this study and discuss the properties of our 500\,$\mu$m risers, followed by the conclusions and summary in Section \ref{sec: conclusions}. Throughout this work, we adopt the standard flat $\Lambda$CDM cosmology: $\Omega_{m} = 0.3$, $\Omega_{\Lambda}= 0.7$, and $H_{0} = 70$\,km\,s$^{-1}$\,Mpc$^{-1}$.

\begin{table*}
    \centering
    \begin{tabular}{cccccccccc}
    \hline\hline \\ [-2.0ex]
        Source & RA & Dec & $S_{250\mu m}$ & $S_{350\mu m}$ & $S_{500\mu m}$ & $S_{850\mu m}$ & $S_{870\mu m}$ & $S_{1.4mm}$  \\ 
         & [J2000] & [J2000] & [mJy] & [mJy] & [mJy] & [mJy] & [mJy] & [mJy] \\ \\ [-2.0ex]
         \hline\hline \\ [-2.0ex]
         Bootes15 & 14:40:09.66 & $+$34:37:55.70 & $51.2 \pm 6.0$ & $68.3 \pm 7.0$ & $66.1 \pm 8.0$ & $2.6 \pm 1.5$ & $3.2 \pm 0.4$ & $-$ \\
         Bootes24 & 14:36:21.30 & $+$33:02:29.00 & $28.7 \pm 6.0$ & $49.7 \pm 7.0$ & $55.8 \pm 8.0$ & $11.3 \pm 2.6$ & $6.8 \pm 0.6$ & $-$ \\
         Bootes27 & 14:38:45.14 & $+$33:22:31.80 & $37.8 \pm 6.0$ & $44.2 \pm 7.0$ & $52.2 \pm 8.0$ & $10.1 \pm 1.6$ & $3.6 \pm 0.6^{\dagger}$ & $-$ \\
         XMM-M5 & 02:18:56.74 & $-$04:35:44.90 & $26.6 \pm 6.0$ & $44.0 \pm 7.0$ & $46.1 \pm 8.0$ & $8.1 \pm 3.3$ & $-$ & $1.3 \pm 0.3$ \\ \\ [-2.0ex]
         \hline\hline \\ [-2.0ex]
         \multicolumn{9}{l}{$^{\dagger}$ Only one of the resolved SMA sources is coincident with the SCUBA-2 contours for Bootes27, but both are coincident } \\ 
         \multicolumn{9}{l}{\hspace{0.12cm} with the \textit{Herschel} SPIRE source.}\\ 
    \end{tabular}
    \caption{Positions and integrated FIR/sub-mm photometry for the four 500\,$\mu$m risers. \textit{Herschel} SPIRE flux densities are taken from G20. The integrated 870\,$\mu$m flux density is calculated by summing the SMA flux densities (corrected for primary beam response) of all sources associated with each 500\,$\mu$m riser in Table \ref{tab: sma}.}
    \label{table: integrated_fir}
\end{table*}

\section{Sample Selection}
\label{sec: sample}

A full description of the selection of the original sample of 500$\mu$m risers is provided in G20. Here we briefly outline the main points of this selection. G20 select a heterogenous sample of thirty-four 500$\mu$m risers based only on their \textit{Herschel} SPIRE flux densities and colours ($S_{250} < S_{350} < S_{500}$) from HerMES. The sample has an average 500$\mu$m flux density of $ 67 \pm 29$\,mJy and all sources are detected to $>4\sigma$. G20 note that recent refinements to the data reduction process means that the flux densities of the sources in the sample vary somewhat in the most up-to-date maps and catalogues, such that $\sim 7$ of the original 34 sources cannot now be strictly defined as 500$\mu$m risers. These sources are still included in the G20 sample as flux boosting is known to introduce some variation into the \textit{Herschel} colours of 500$\mu$m risers, and typical star-forming SED shapes indicate that these sources could still reside at high redshift. G20 used the ALESS \citep{2015ApJ...806..110D} average SED at redshifts between 4 and 6 and normalised to a 500\,$\mu$m flux density of 60\,mJy to estimate that, given their observed sub-mm flux densities from HerMES and assuming that a single source is responsible for the total \textit{Herschel} flux density, these sources should have flux densities of $\sim 24-38$\,mJy at 345\,GHz. However, despite reaching detection thresholds of $\sim 7 - 10$\,mJy, twelve of these maps showed no detections. Assuming in the most conservative case that any multiple systems are comprised of sources with similar flux densities, G20 predicted that these 500\,$\mu$m risers must break up into $\geq 3$ individual sources, and are therefore potential indications of forming compact cluster cores at $z > 4$ similar to those discovered by \cite{2018ApJ...856...72O} and \cite{2018Natur.556..469M}. 

\begin{figure}
\hspace{-0.2cm}
\includegraphics[width=0.45\textwidth]{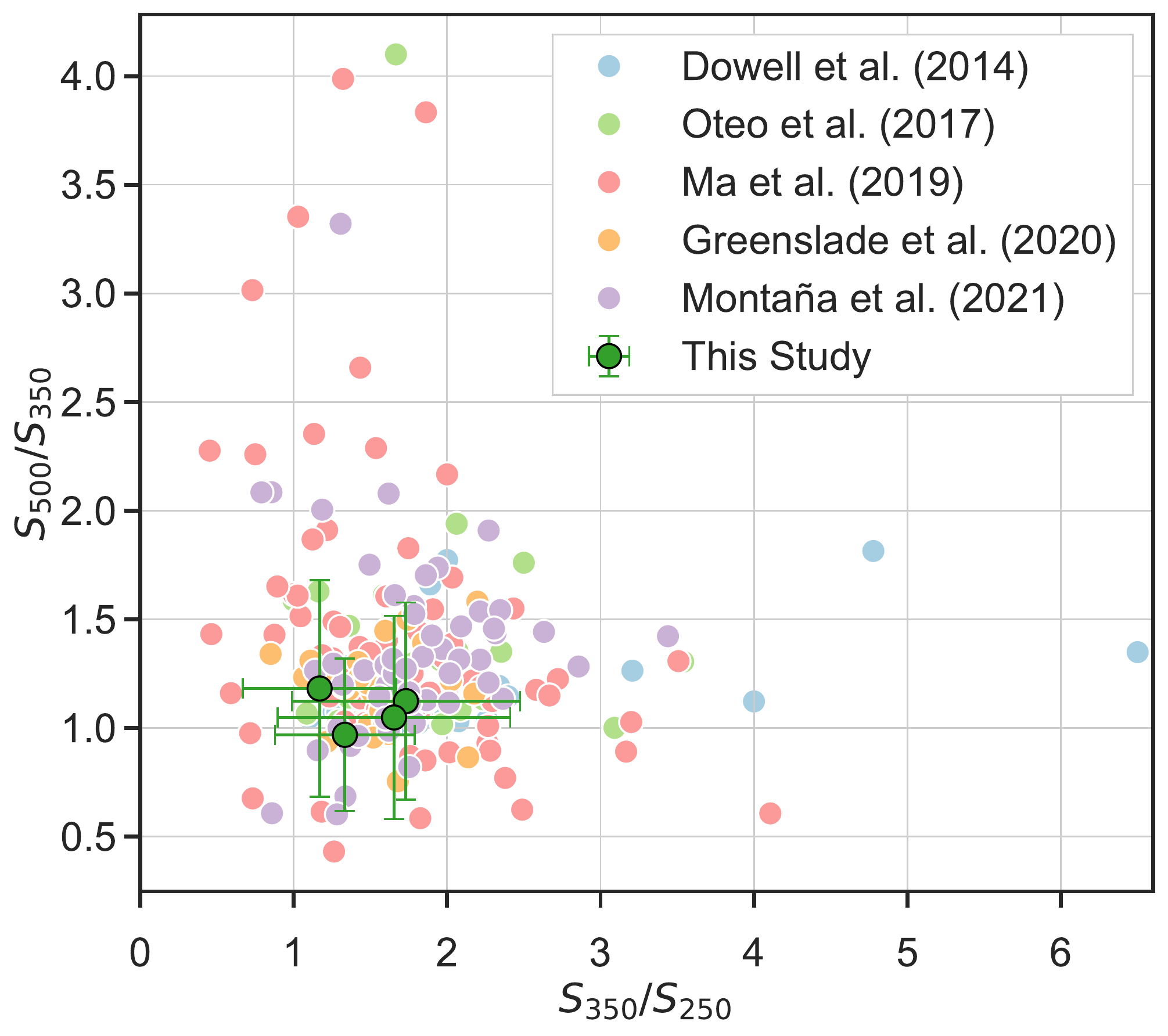}
\caption{\textit{Herschel} SPIRE colours for our 500\,$\mu$m risers alongside those from a selection of studies in the literature. Note that the \citet{2019ApJS..244...30M} and \citet{2021MNRAS.505.5260M} colours are based on deblended \textit{Herschel} photometry using high resolution follow-up observations and so contain a number of sources that are no longer strictly 500\,$\mu$m risers. We find that our 500\,$\mu$m risers are towards the less extreme end of the global 500\,$\mu$m riser population.}
\label{fig: sample_colours}
\end{figure}

We selected the three brightest sources at 500\,$\mu$m from the twelve sources in the G20 sample without SMA counterparts (Bootes15, Bootes24 and Bootes27) for multi-wavelength follow-up observations. An additional 500\,$\mu$m riser (XMM-M5) was observed as part of a separate SMA project, but will also be presented in this paper. The positions and \textit{Herschel} SPIRE photometry for these sources are presented in Table \ref{table: integrated_fir}, and in Figure \ref{fig: sample_colours} we show their \textit{Herschel} SPIRE colours. We find that our 500\,$\mu$m risers are quite typical of the majority of the global 500\,$\mu$m riser population, although potentially residing towards
the less extreme end. 

\section{Observations and Data Reduction}
\label{sec: bootes_observations}

\subsection{SMA Observations}
\label{sec: SMA_obs}

The SMA observations were taken in three tracks between 31$^{\text{st}}$ January and 26$^{\text{th}}$ February 2020 as part of the SMA program 2019B-S016 (PI: J. Cairns). The conditions for these observations were exquisite, with $\tau$ values of $\sim 0.02 - 0.04$. We used the SMA in its COM configuration with 7 antennae, resulting in a beamsize of $\sim 1.7^{\prime\prime} \times 2.1^{\prime\prime}$. The SMA uses two receivers, each with two sidebands of 8\,GHz separated by a gap of 8\,GHz, and with each sideband separated into four chunks. We tuned these two receivers to 337\,GHz and 345\,GHz respectively to produce 32\,GHz of continuous bandwidth for our observations. 

\begin{figure*}
\centering
\includegraphics[width=0.9\textwidth]{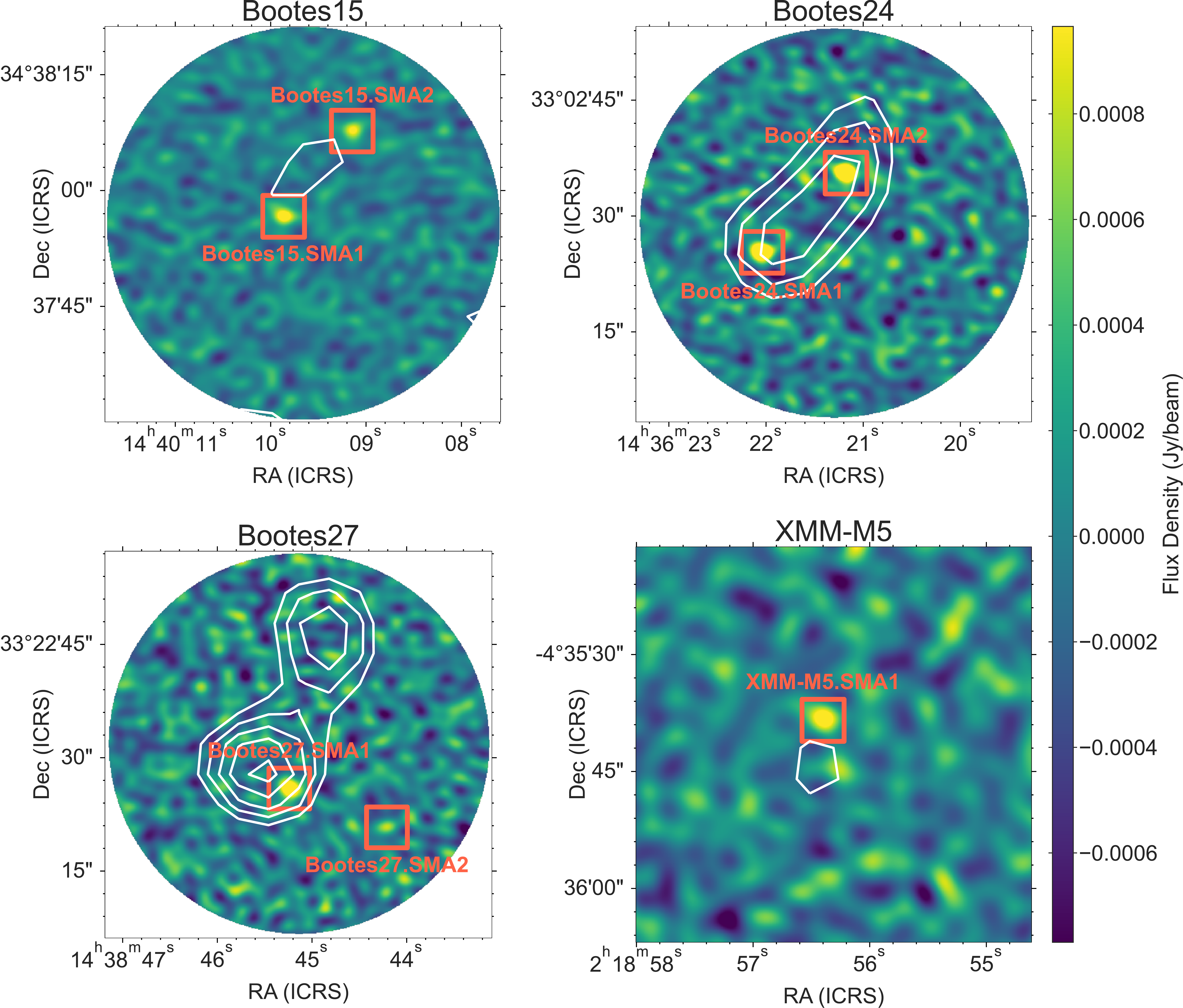}
\caption{SMA maps overlaid with SCUBA-2 850\,$\mu$m contours from $3\sigma$ to $7\sigma$ significance for the four 500\,$\mu$m risers. Bootes15, Bootes24 and Bootes27 each break up into two individual sources in the SMA maps, while XMM-M5 resolves into one faint source. Bootes24 and Bootes27 are well detected by SCUBA-2, while Bootes15 and XMM-M5 are marginally detected to $\sim 3\sigma$ significance.} 
\label{fig: sma_maps}
\end{figure*}

We reduced our SMA maps using the standard IDL-based SMA data reduction package MIR\footnote{\url{https://github.com/qi-molecules/sma-mir}}. Firstly, we manually inspected the data, flagging any regions that contained phase jumps and fixing any channels that contained significant noise spikes, before applying the system temperature correction. The passband calibration was completed using the sources 3c454.3, 3c279, 0927+390 or 3c84, while for flux calibration, we used either 0854+201 or Callisto. Gain calibration on the SMA is carried out by periodically observing nearby brights quasars with known flux densities alternately to the desired source. For these purposes, we used the quasars 1506+426 and 3c345. The error for this calibration process is estimated to be $\sim 10\%$ \citep[see e.g.][]{2018A&A...612A..54L}.  

We then used the Common Astronomy Software Applications package\footnote{\url{https://casa.nrao.edu/index.shtml}} \cite[CASA:][]{2007ASPC..376..127M} to complete the imaging. Following the calibration process in MIR, each chunk of each sideband on each receiver is exported into CASA, before being concatenated together to produce one large visibility dataset. We manually flagged the $\sim 300$ channels on the edge of each chunk in order to avoid including poor data in the final image, inspecting the result to ensure that we had flagged enough of the edge channels and that there were no spikes or strong lines in the data. We generated the continuum by averaging all of the remaining spectral channels together. In order to clean the SMA maps, we first identified the positions of likely sources based on the dirty maps. We followed G20 who found that a detection threshold of $3.75\sigma$ (where $\sigma$ represents the global noise in the dirty map) without any corrections for the response of the primary beam produced a good balance between minimising the number of fake sources extracted and minimising the signal-to-noise threshold for extraction. We therefore found all sources above this $3.75\sigma$ detection threshold in the dirty maps and considered them to be real sources. We used the CASA command \code{tclean} to complete interactive CLEANing of our SMA maps. We selected natural weighting which maximises the signal-to-noise and used $<100$ CLEAN iterations for each map, placing CLEAN windows in the images to preferentially select flux from the positions of the $>3.75\sigma$ detections. The resulting CLEANed maps reached 1$\sigma$ RMS values of 0.17, 0.29 and 0.27\,mJy/beam for Bootes15, Bootes24 and Bootes27 respectively, around an order of magnitude deeper than the maps obtained by G20. 

XMM-M5 was observed as part of the SMA project 2019A-S004 (PI: D. L. Clements). The observations and subsequent data reduction will be discussed in a forthcoming paper (Clements et al. in prep) and we simply highlight the important features here. The SMA was used in its COM configuration with the two receivers tuned to 198\,GHz and 206\,GHz respectively, resulting in a beamsize of $\sim 3.6^{\prime\prime} \times 3.2^{\prime\prime}$. The reduced SMA map for XMM-M5 has a $1\sigma$ RMS sensitivity of $\sim 0.28$\,mJy/beam. The CLEANed SMA maps for all four sources are shown in Figure \ref{fig: sma_maps}.

\begin{table*}
    \centering
    \begin{tabular}{ccccc}
    \hline\hline \\ [-2.0ex]
        Source & RA  & Dec & $S_{870\mu m}$ & Primary Beam \\
        & [J2000] & [J2000] & [mJy] & Response \\ \\ [-2.0ex]
        \hline\hline \\ [-2.0ex]
        Bootes15.SMA1 & 14:40:09.86 & $+$34:37:56.65
        & $1.4 \pm 0.2$ & 0.98\\ 
        Bootes15.SMA2 & 14:40:09.13 & $+$34:38:07.84
        & $1.8 \pm 0.3$ & 0.60 \\ [+1.0ex] \hline \\ [-2.0ex]
        Bootes24.SMA1 & 14:36:22.04 & $+$33:02:25.30
        & $3.4 \pm 0.5$ & 0.76\\
        Bootes24.SMA2 & 14:36:21.17 & $+$33:02:35.53
        & $3.4 \pm 0.4$ & 0.88\\ [+1.0ex] \hline \\ [-2.0ex]
        Bootes27.SMA1 & 14:38:45.24 & $+$33:22:25.92
        & $1.9 \pm 0.3$ & 0.91\\
        Bootes27.SMA2 & 14:38:44.20 & $+$33:22:20.79
        & $1.7 \pm 0.5$ & 0.49\\[+1.0ex] \hline \\ [-2.0ex]
        XMM-M5.SMA1 & 02:18:56.39 & $-$04:35:38.32
        & $1.3 \pm 0.3^{\dagger}$ & 0.99 \\ \\ [-2.0ex]
        \hline\hline \\ [-2.0ex]
        \multicolumn{4}{l}{$^{\dagger}$ SMA observations tuned to 210\,GHz rather than 345\,GHz.}\\ 
    \end{tabular}
    \caption{Positions, corrected SMA flux densities and primary beam response values for the resolved SMA sources associated with each 500\,$\mu$m riser.}
    \label{tab: sma}
\end{table*}

\subsection{SCUBA-2 Observations}
\label{sec: SCUBA_obs}

The JCMT/SCUBA-2 450\,$\mu$m and 850\,$\mu$m observations for Bootes15, Bootes24 and Bootes27 were taken between 1$^{\text{st}}$ December 2020 and 20$^{\text{th}}$ January 2021 as part of the SCUBA-2 program M20BP036 (PI: J. Cairns). Each source was observed with several pointings using the CV Daisy mode, with a total of $\sim 2.7$ hours on Bootes15, $\sim 1.4$ hours on Bootes24 and $\sim 4.1$ hours on Bootes27. Weather conditions were typically between Band 2 and Band 3, with $\tau_{225}$ values in the range $0.08 - 0.14$. This allowed us to reach 1$\sigma$ RMS sensitivities of 2.2, 2.8 and 1.7\,mJy/beam in our SCUBA-2 850\,$\mu$m maps of Bootes15, Bootes24 and Bootes27 respectively. We reduced our observations using the Sub-Millimetre User Reduction Facility (\texttt{SMURF}) package \citep{2013MNRAS.430.2545C} and the \texttt{ORAC-DR} data reduction pipeline \citep{1999ASPC..172...11E}. 

For the first reduction we completed each step of the data reduction process manually following \citet{2019MNRAS.490.3840C}. We first used the \texttt{MAKEMAP} command on the individual scans from SCUBA-2, with the \texttt{METHOD} parameter set to the default \texttt{ITERATE} method, which uses an iterative technique to fit a number of models for noise and instrumental behaviour. We then co-added the individual scans using the \texttt{PICARD} recipe \texttt{MOSAIC\_JCMT\_IMAGES} to produce a single map. This step also removes a number of contaminant signals. For this first reduction, we assumed that the individual sources detected in the SMA images would be contained within the much larger SCUBA-2 beam, and so would appear as a single point-like source. We applied a matched filter using the \texttt{PICARD} recipe \texttt{SCUBA2\_MATCHED\_FILTER}, which subtracts the background by convolving the
maps and the PSF with a 30$^{\prime\prime}$ FWHM Gaussian kernel, before convolving the maps with the PSF to produce the matched-filtered signal map. This
process gives an effective beam FWHM of 14.6$^{\prime\prime}$ surrounded by a shallow negative ring, and is commonly used for finding sources with angular scales of a similar size or smaller to the beamsize of the SCUBA-2 instrument. The signal maps produced using this method are in units of pW, and so must be calibrated using a Flux Correction Factor (FCF). We used the standard FCF value\footnote{\url{https://www.eaobservatory.org/jcmt/instrumentation/continuum/scuba-2/calibration/}} of $537 \pm 43$\,Jy\,beam$^{-1}$\,pW$^{-1}$. 

For our second data reduction method, we made use of the \texttt{ORAC-DR} data reduction pipeline. We first used the \texttt{REDUCE\_SCAN\_FAINT\_POINT\_SOURCES} recipe which employs a similar method to our manual data reduction process above. Raw data are passed to the map maker to produce an image calibrated in mJy/beam. The pipeline then estimates the RMS noise in the image and calculates the Noise Equivalent Flux Density (NEFD). Once all the individual observations have been processed, the pipeline co-adds individual scans together, which it then convolves with a matched filter to enhance the signal-to-noise ratio of point sources. The RMS noise and NEFD are calculated for this co-added image, and a signal-to-noise map is produced. We find that the pipeline produces essentially identical results to our manual data reduction.

Finally, we completed a third data reduction process using the \texttt{ORAC-DR} pipeline with the \texttt{REDUCE\_SCAN\_EXTENDED\_SOURCES} recipe. Given that the SCUBA-2 beamsize is smaller than that of \textit{Herschel} SPIRE at 250\,$\mu$m but larger than that of the SMA, it is possible that any multiple systems may be partially resolved in the SCUBA-2 850\,$\mu$m maps. If this is the case then flux density estimates should be calculated based on SCUBA-2 maps reduced without the matched filter to avoid missing flux for the partially resolved sources lying outside of the SCUBA-2 beam. The \texttt{REDUCE\_SCAN\_EXTENDED\_SOURCES} recipe passes the raw data for the individual scans to the map maker which processes them to produce a Frame image, which is then calibrated in units of mJy\,arcsec$^{-2}$. Individual scans are then co-added together and the noise properties of this image are calculated. The analysis of the SCUBA-2 maps in the following sections is based on the two pipeline reductions. While the SCUBA-2 instrument simultaneously provides 450\,$\mu$m and 850\,$\mu$m photometry, the noise levels in the 450\,$\mu$m maps are too large to detect our 500\,$\mu$m risers based on their \textit{Herschel} SPIRE flux densities.

We additionally used archival 850\,$\mu$m SCUBA-2 maps for XMM-M5 from the Canadian Astronomy Data Center. XMM-M5 was observed as part of the SCUBA-2 Large eXtragalactic Survey (S2LXS\footnote{\url{https://www.eaobservatory.org/jcmt/science/large-programs/s2lxs/}}: Geach et al. M17BL001) and is present in two overlapping pointings. We re-reduced each of these pointings using the \texttt{ORAC-DR} pipeline as outlined above, and co-added the two pointings to obtain a deeper SCUBA-2 850\,$\mu$m image of XMM-M5. After this reduction process, we find that the RMS noise in an aperture of radius 350$^{\prime\prime}$ centered on the SMA position of XMM-M5 is $\sim3.3$\,mJy/beam. In Figure \ref{fig: sma_maps} we show the SMA maps for the 500\,$\mu$m risers with 850\,$\mu$m contours based on the \texttt{REDUCE\_SCAN\_FAINT\_POINT\_SOURCES} reduced SCUBA-2 maps. The SCUBA-2 maps with the matched filter were chosen for Figure \ref{fig: sma_maps} as this reduction maximises the signal-to-noise ratio in the map.

\subsection{Extracting Flux Densities}
\label{sec: extracting_flux}

We then extracted flux densities from the SMA and SCUBA-2 maps. G20 found that the most robust method for extracting the flux density of point-like SMA sources is to simply extract the peak flux density directly from the CLEANed SMA map. We therefore extracted peak flux densities from the CLEANed SMA maps for each source detected to $>3.75\sigma$ in the dirty SMA maps. These flux densities were then corrected for the primary beam response. The response of the primary beam for the SMA can be described as a Gaussian function with a size determined by the wavelength of the observations\footnote{\url{https://lweb.cfa.harvard.edu/sma/miriad/manuals/SMAuguide/smauserhtml/node130.html}}. We used the \texttt{pbplot} command in the MIRIAD \citep{2011ascl.soft06007S} package to determine the FWHM of the SMA primary beam at 345\,GHz for our three sources in the Bootes field, and at 210\,GHz for XMM-M5. We then found the primary beam response at the position of the peak flux density value for each source, and divided the extracted flux density by this value, propagating the errors accordingly. In Table \ref{tab: sma} we present the corrected SMA photometry alongside the corresponding correction factor for the primary beam response, where the uncertainties include the 10\% calibration error added in quadrature with the RMS noise in the SMA map. We also extracted flux densities from the dirty SMA maps using a similar method, finding that they differ by no more than $\sim 3\%$ from those extracted from the CLEANed maps.

For the SCUBA-2 850\,$\mu$m maps, we extracted two separate flux densities, one for each of the two pipeline reductions. For the 850\,$\mu$m map reduced using the \texttt{REDUCE\_SCAN\_FAINT\_POINT\_SOURCES} recipe, we located the highest signal-to-noise pixel associated with the detection and extracted the corresponding flux density. For the 850\,$\mu$m map reduced using the \texttt{REDUCE\_SCAN\_EXTENDED\_SOURCES} recipe, we laid down a series of apertures in the image with increasing radii from 0$^{\prime\prime}$ to 10$^{\prime\prime}$, and calculated the sum of the pixel values within these apertures. This produces a flux density in units of mJy\,arcsec$^{-2}$, which we then multiplied by 16 in order to convert to mJy (based on the 4$^{\prime\prime}$ pixel size of the SCUBA-2 850\,$\mu$m images). For the final flux density measurement, we selected the aperture with the largest flux density value, as this aperture should contain the maximum amount of flux density from the source without including too much contaminating background (we also manually checked each final aperture to make sure that it was a reasonable size). 

SCUBA-2 850\,$\mu$m photometry must be corrected for flux boosting. \cite{2017MNRAS.465.1789G} investigated the effects of flux boosting in the SCUBA-2 Cosmology Legacy Survey (S2CLS), finding that the level of flux boosting is consistent across the whole survey and well described by the power law
\begin{eqnarray}
    B = 1 + 0.2 \left( \frac{\text{SNR}}{5} \right)^{-2.3}
\end{eqnarray}

where $B$ is the flux boosting factor and SNR is the signal-to-noise ratio of the detection. We estimated errors for our SCUBA-2 flux density measurements based on the combination of instrumental noise, confusion noise and the typically assumed 5\% calibration error.

\subsection{Ancillary Data}
\label{sec: ancillary_data}

Due to their position towards the edge of the field, our three 500\,$\mu$m risers in Bootes only benefit from sporadic ancillary data. All three are covered by the DESI Legacy Imaging Surveys\footnote{\url{http://legacysurvey.org/}} which combines the Dark Energy Camera Legacy Survey, the Beijing–Arizona Sky Survey, and the Mayall $z$-band Legacy Survey, mapping $\sim 14,000$\,deg$^{2}$ of the sky using the Blanco telescope at the Cerro Tololo Inter-American Observatory, as well as the Mayall and 2.3\,m Bart Bok Telescopes at the KPNO down to AB magnitudes of 23.72, 22.87 and 22.29 in the $g$, $r$ and $z$ bands respectively \citep{2019AJ....157..168D}. There is a tenuous detection in $g$, $r$ and $z$ for one of the resolved objects associated with Bootes15, but the remainder are undetected in these images.

The Infrared Bootes Imaging Survey \cite[IBIS:][]{2010AAS...21641513G} provides deep NIR images and catalogues in the $J$, $H$ and $K_{s}$ bands, down to limiting AB magnitudes of 22.0, 21.5, and 20.8  respectively, and we extract 3.4, 4.6, 12, and 22\,$\mu$m \textit{Wide-Field Infrared Survey Explorer} \cite[\textit{WISE}:][]{2010AJ....140.1868W} images and catalogues from the ALLWISE database, which combines the \textit{WISE} cryogenic and NEOWISE \citep{2011ApJ...731...53M} post-cryogenic phases. Additionally, the \textit{Spitzer} Deep Wide-Field Survey \cite[SDWFS:][]{2009ApJ...701..428A} provides deep 3.6, 4.5, 5.8 and 8.0\,$\mu$m imaging for 10\,deg$^{2}$ in the Bootes field. Bootes24 resides just within the footprint of SDWFS, although it only benefits from data at 4.5 and 8.0\,$\mu$m. For Bootes15 and Bootes27, we rely on archival IRAC imaging. Bootes15 is covered at 3.6 and 4.5\,$\mu$m as part of \textit{Spitzer} Program 80156 (PI: A. Cooray) aiming to study lensed sub-mm galaxies from the \textit{Herschel}-ATLAS \cite[H-ATLAS:][]{2010PASP..122..499E} and HerMES surveys. Additionally, Bootes27 is covered at 4.5 and 8.0\,$\mu$m as part of \textit{Spitzer} Program 30134 (PI: G. Fazio) aiming to study a statistically complete sample of star-forming dwarf galaxies. All three 500\,$\mu$m risers remain undetected in the IBIS catalogue, and only one of the resolved sources associated with Bootes15 is detected in the \textit{WISE} images (although this \textit{WISE} detection appears to be a blend of two IRAC sources). The two sources associated with Bootes15 are both detected in the two IRAC bands. Those associated with Bootes24 are well detected at 4.5\,$\mu$m but not at 8.0\,$\mu$m. There are numerous bright 8.0\,$\mu$m IRAC sources associated with Bootes27, and a more comprehensive, multi-wavelength analysis of this source will be presented in a forthcoming paper (Cairns et al. in prep). Given the sparse ancillary data in the optical and NIR for our Bootes sources, it is not discussed further in this paper.  

\begin{figure*}
\centering
\includegraphics[width=0.9\textwidth]{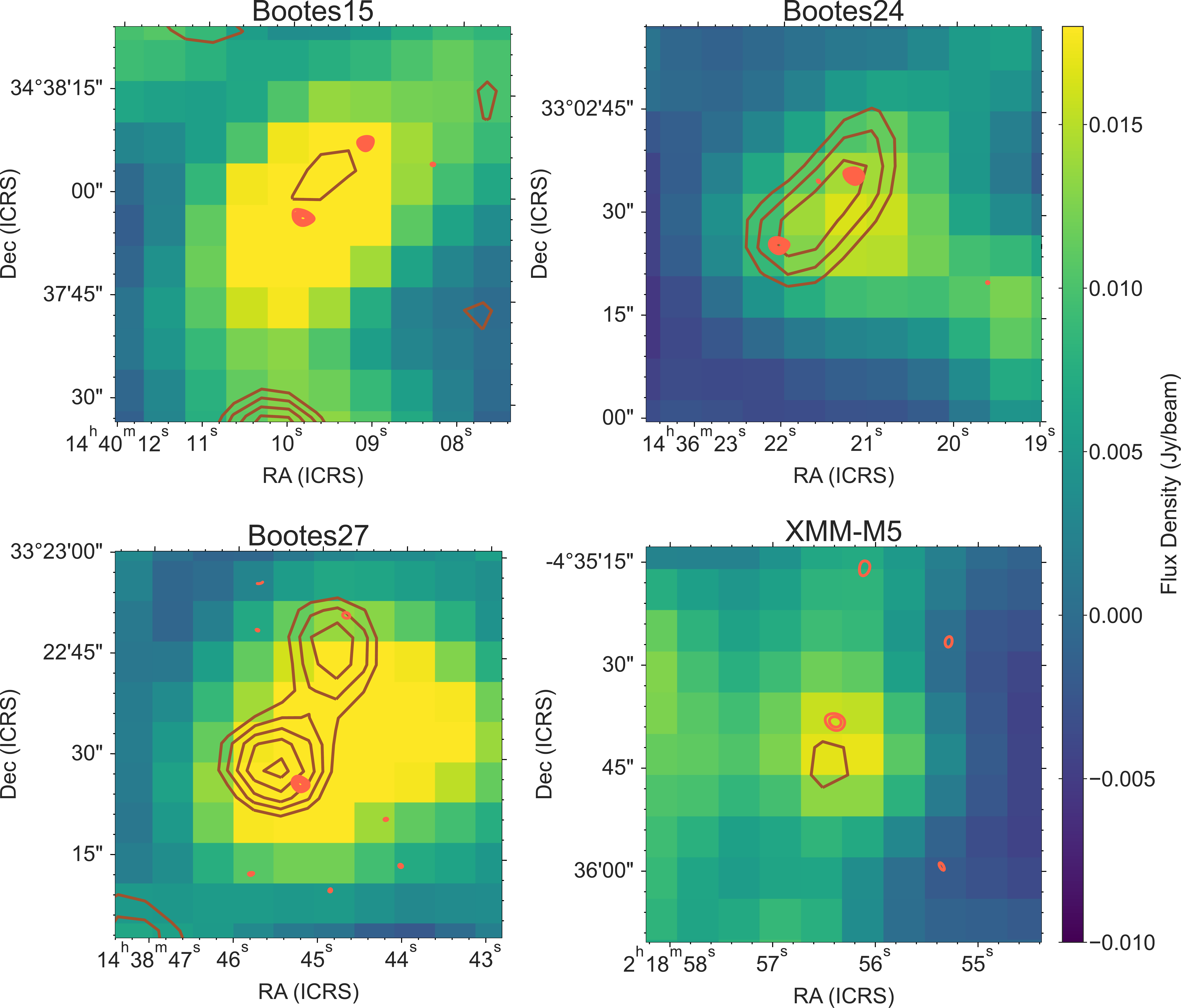}
\caption{\textit{Herschel} SPIRE 250\,$\mu$m maps with SCUBA-2 850\,$\mu$m (brown) and SMA (red) contours from $3\sigma$ to $7\sigma$ significance for the four 500\,$\mu$m risers. Bootes15, Bootes24 and Bootes27 each break up into two individual sources in the SMA maps, while XMM-M5 resolves into one faint source. Bootes24 and Bootes27 are well detected by SCUBA-2, while Bootes15 and XMM-M5 are marginally detected to $\sim 3\sigma$ significance. Bootes27 shows two partially resolved SCUBA-2 sources associated with the bright \textit{Herschel} source.}
\label{fig: herschel_map}
\end{figure*}

By contrast, XMM-M5 benefits from a wealth of multi-wavelength data, including `forced photometry' at optical and NIR wavelengths from \cite{2017ApJS..230....9N}. This includes photometry in the $u$, $g$, $r$, $i$ and $z$ bands from the Canada-France-Hawaii Telescope Legacy Survey \cite[CFHTLS:][]{2012AJ....143...38G}, in the $Z$, $Y$, $J$, $H$ and $K_{s}$ bands from the VISTA Deep Extragalactic Observations \cite[VIDEO:][]{2013MNRAS.428.1281J} survey, and in the 3.6 and 4.5\,$\mu$m IRAC bands from the \textit{Spitzer} Extragalactic Representative Volume Survey \cite[SERVS:][]{2012PASP..124..714M}. The reader is directed to \cite{2017ApJS..230....9N} for a description of how this photometry was extracted. XMM-M5 is also included in the Subaru XMM Deep Survey \citep[SXDS:][]{2008ApJS..176....1F} which covers five broadband filters reaching limiting AB magnitudes of $B = 28.4$, $V = 27.8$, $R_{c} = 27.7$, $i^{\prime} = 27.7$, and $z^{\prime} = 26.6$, as well as the Hyper Suprime-Cam Subaru Strategic Program \citep[HSC SSP:][]{2018PASJ...70S...4A} Survey which, in its `Deep' layer, reaches limiting AB magnitudes of 27.5, 27.1, 26.8, 26.3 and 25.3 in the $g$, $r$, $i$, $z$ and $y$ bands respectively. A more comprehensive, multi-wavelength analysis of XMM-M5 will be presented in Clements et al. (in prep).

\section{Results}
\label{sec: results}

\subsection{SMA and SCUBA-2 Detections}
\label{sec: SMA_SCUBA_detection}

In Figure \ref{fig: sma_maps}, we present the SMA maps for the four 500\,$\mu$m risers, overlaid with SCUBA-2 850\,$\mu$m contours. In Figure \ref{fig: herschel_map}, we present a similar plot showing the 250\,$\mu$m \textit{Herschel} SPIRE map overlaid with both SCUBA-2 850\,$\mu$m and SMA contours. We find that Bootes15 and Bootes24 each break up into two bright sources, both of which are detected to $>5\sigma$ in both the dirty and CLEANed SMA maps. Bootes27 also breaks up into two resolved sources, but while the brighter source is similarly detected to $>5\sigma$, the fainter source is a more marginal detection at $\sim3.9\sigma$ in the dirty map and $\sim3.1\sigma$ in the CLEANed map. Moreover, this fainter source is offset by $\sim 10^{\prime\prime}$ from the edge of the observed SCUBA-2 $3\sigma$ emission region, and so, while it may contribute to the integrated \textit{Herschel} SPIRE flux density, it is unlikely to contribute to the observed 850\,$\mu$m flux density. After correcting for the primary beam, we find that this fainter source has a flux density of $1.7 \pm 0.5$\,mJy and, as a result, we would not expect to detect it given the 1.7\,mJy/beam RMS noise in the SCUBA-2 map and the $\sim 10^{\prime\prime}$ offset from the main emission region. There is another marginal $\sim3.3\sigma$ detection in the dirty SMA map coinciding with the Northern component of the SCUBA-2 contours (Figure \ref{fig: herschel_map}), but as this falls below our $3.75\sigma$ detection threshold we do not consider this detection robust. The SMA map for XMM-M5 shows a single $\sim 4.7\sigma$ peak coincident with the bright \textit{Herschel} 500\,$\mu$m riser (Figure \ref{fig: herschel_map}).

The SCUBA-2 850\,$\mu$m contours in Figures \ref{fig: sma_maps} and \ref{fig: herschel_map} demonstrate that XMM-M5 is marginally detected as a point-like source with a signal-to-noise ratio (SNR) of 3.65. Bootes15 is also marginally detected with a signal-to-noise ratio of 3.06, but is likely to be extended as the SCUBA-2 flux density peaks between the two SMA detections and is elongated along the direction of their separation. By comparison, Bootes24 and Bootes27 are both well detected to $>5\sigma$ and appear partially resolved in the SCUBA-2 contours. Bootes27 in particular is clearly comprised of two partially resolved sources, one of which has no apparent SMA cross-match. After correcting for flux boosting we obtain flux densities of $2.9 \pm 1.8$\,mJy, $9.7 \pm 2.2$\,mJy, $6.9 \pm 1.3$\,mJy and $8.1 \pm 3.3$\,mJy for Bootes15, Bootes24, Bootes27 and XMM-M5 respectively based on the 850\,$\mu$m maps reduced using the \texttt{REDUCE\_SCAN\_FAINT\_POINT\_SOURCES} recipe. Using the 850\,$\mu$m maps reduced using the  \texttt{REDUCE\_SCAN\_EXTENDED\_SOURCES} recipe, we obtain flux densities of $2.6 \pm 1.5$\,mJy, $11.3 \pm 2.6$\,mJy and $10.1 \pm 1.6$\,mJy for Bootes15, Bootes24 and Bootes27 respectively, while no reliable flux density could be extracted for XMM-M5 using this reduction pipeline. The flux density estimates from the two pipeline reduction methods agree within the errors for all sources except for Bootes27, where the significantly larger flux density based on the \texttt{REDUCE\_SCAN\_EXTENDED\_SOURCES} recipe reflects the more extended nature of the object. For the analysis in this paper, we will use the flux density from the \texttt{REDUCE\_SCAN\_FAINT\_POINT\_SOURCES} reduction for XMM-M5 which appears as a point source in the SCUBA-2 maps, and the flux densities from the \texttt{REDUCE\_SCAN\_EXTENDED\_SOURCES} reduction for the three sources in the Bootes field which appear as extended sources. We additionally note that, while Bootes15 and XMM-M5 are detected to $>3\sigma$ in the SCUBA-2 850\,$\mu$m maps, after including the instrumental, confusion and calibration uncertainties, the photometry is constrained to an accuracy of $\sim1.7\sigma$ and $\sim2.5\sigma$ respectively.

\begin{figure}
\hspace{-0.2cm}
\includegraphics[width=0.45\textwidth]{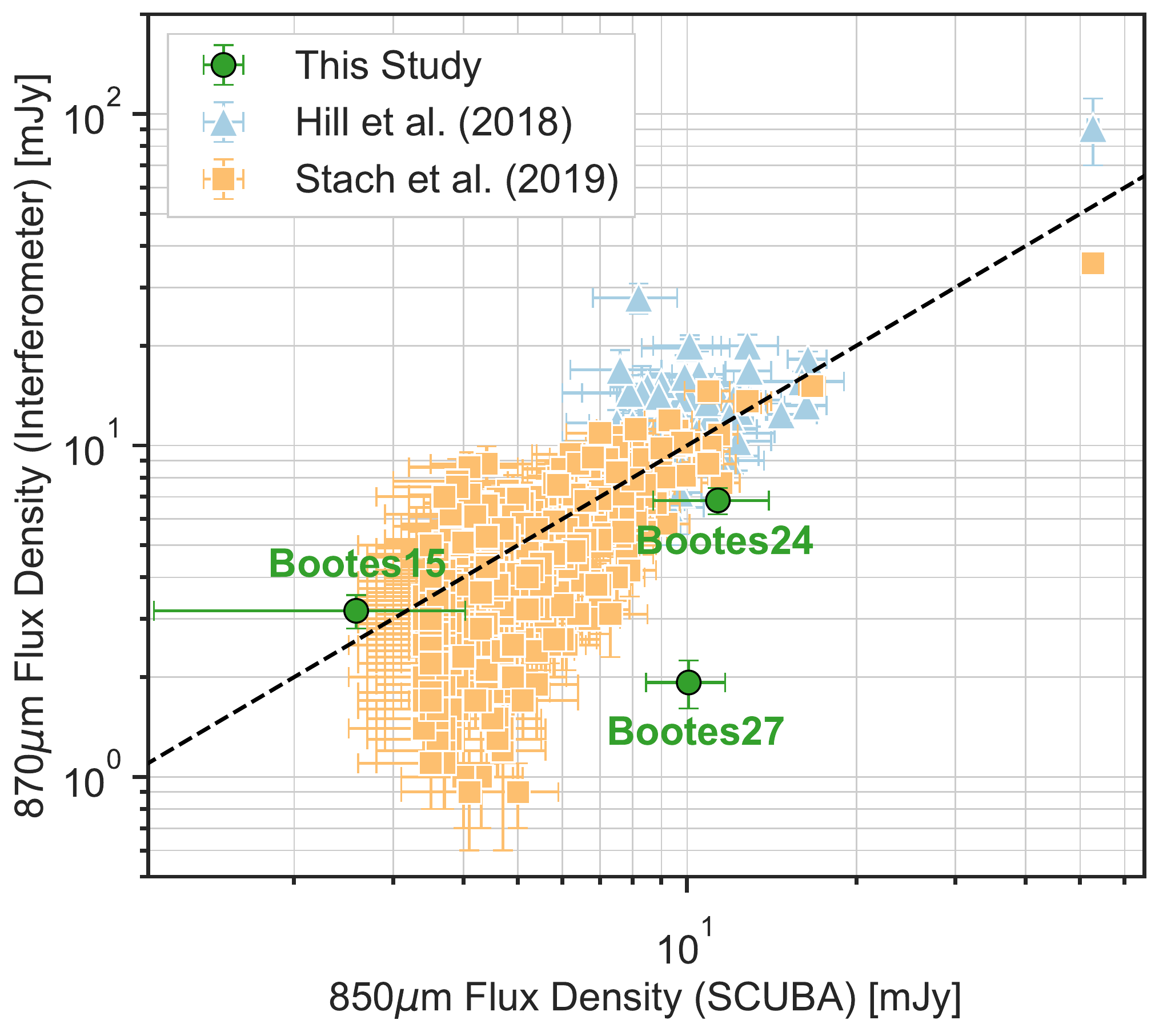}
\caption{The combined 870\,$\mu$m flux density for the resolved sources in either SMA or ALMA maps as a function of their integrated 850\,$\mu$m flux density from SCUBA-2. Orange squares and blue triangles show the \citet{2019MNRAS.487.4648S} and \citet{2018MNRAS.477.2042H} samples respectively, while the green points show our sample of 500\,$\mu$m risers. The black dashed line shows where the two flux density measurements are equal. We do not include XMM-M5 in this plot as the SMA observations are tuned to a shorter wavelength (210\,GHz rather than 345\,GHz).}
\label{fig: interferometer_vs_single-dish}
\end{figure}

In Table \ref{table: integrated_fir}, we present the integrated FIR/sub-mm photometry for the four 500\,$\mu$m risers, where we have estimated the total SMA flux density for each source by adding together the flux densities of the resolved sources in the SMA continuum maps from Table \ref{tab: sma}. We find that, despite being observed at similar wavelengths, the integrated 850\,$\mu$m flux densities from SCUBA-2 are larger than the combined flux densities of the detected SMA sources at 870\,$\mu$m for Bootes24 and Bootes27 (note that XMM-M5 is observed at a lower frequency and so we exclude it from this analysis). This disparity is particularly evident in Bootes27 - if we exclude the fainter SMA source that does not appear to contribute to the observed SCUBA-2 flux density, we measure integrated flux densities of $10.1 \pm 1.6$\,mJy and $1.9 \pm 0.3$\,mJy from the SCUBA-2 and SMA maps respectively, indicating a $>4\sigma$ disparity between the two measurements. We investigate this discrepancy by plotting the SCUBA-2 850\,$\mu$m flux density against the SMA continuum flux density at $\sim 870$\,$\mu$m for each of the 500\,$\mu$m risers alongside similar studies which obtained both interferometric and single-dish sub-mm photometry, where for Bootes27 we include only the SMA flux density for the resolved source coincident with the SCUBA-2 emission. \cite{2018MNRAS.477.2042H} targeted 70 of the brightest 850\,$\mu$m sources from S2CLS down to a limiting flux density of $S_{850} \sim 8$\,mJy with high-resolution SMA follow-up observations at $860$\,$\mu$m. Similarly, \cite{2019MNRAS.487.4648S} present the ALMA survey of the SCUBA-2 Cosmology Legacy Survey UKIDSS/UDS field (AS2UDS), providing high resolution, ALMA Band 7 follow-up observations of SCUBA-2 850\,$\mu$m sources in the UKIDSS/UDS field. We combined the SMA, ALMA and SCUBA-2 photometry from these surveys and, for any SCUBA-2 sources which break up into multiple components in the higher resolution interferometric observations, we add the flux densities of the resolved sources together in order to compare with our own photometry. The result of this comparison is shown in Figure \ref{fig: interferometer_vs_single-dish}. We find that the SMA and SCUBA-2 flux densities for Bootes15 are consistent.  Bootes24 is $\sim1.7$ times fainter in the SMA maps than the SCUBA-2 maps and lies towards the edge of the regions probed by the \cite{2018MNRAS.477.2042H} and \cite{2019MNRAS.487.4648S} samples in Figure \ref{fig: interferometer_vs_single-dish}, but the two values are within $3\sigma$ of each other. For Bootes27, the single resolved SMA source coincident with the SCUBA-2 contours is roughly five times fainter than the integrated flux density from the SCUBA-2 maps, placing it well outside of the region probed by the \cite{2018MNRAS.477.2042H} and \cite{2019MNRAS.487.4648S} samples. We interpret this disparity as evidence of additional faint sources residing below the SMA detection limit, but contributing to the integrated flux density at 850\,$\mu$m.

\begin{figure}
\hspace{-0.2cm}
\includegraphics[width=0.45\textwidth]{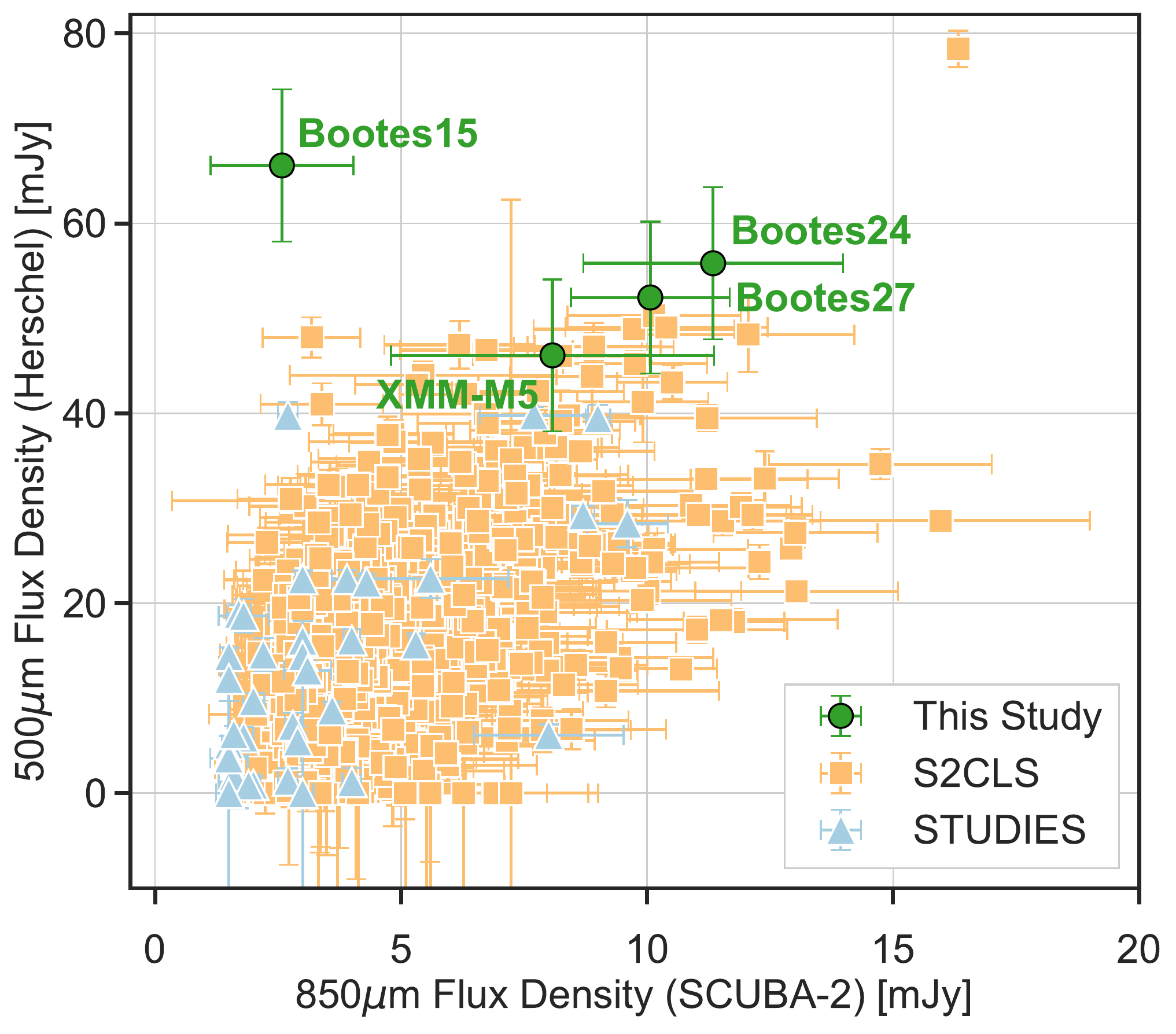}
\caption{SCUBA-2 850\,$\mu$m flux density vs. \textit{Herschel} SPIRE 500\,$\mu$m flux density for our 500\,$\mu$m risers (green points) as well as for the S2CLS (orange squares) and STUDIES (blue triangles) surveys. Bootes15 is significantly fainter than expected at 850\,$\mu$m based on its 500\,$\mu$m flux density.}
\label{fig: 850v500}
\end{figure}

Moreover, we find that the integrated SCUBA-2 850\,$\mu$m flux densities of our 500\,$\mu$m risers are somewhat lower than we might expect based on their \textit{Herschel} SPIRE colours. In Figure \ref{fig: 850v500} we compare the integrated 500\,$\mu$m and 850\,$\mu$m flux densities of our 500\,$\mu$m risers, alongside the S2CLS and STUDIES \cite[SCUBA-2 Ultra Deep Imaging EAO Survey:][]{2017ApJ...850...37W} samples. While the FIR/sub-mm photometry for Bootes24, Bootes27 and XMM-M5 is reasonably consistent with the brighter end of S2CLS and STUDIES, Bootes15 is significantly fainter at 850\,$\mu$m than sources with similar 500\,$\mu$m flux densities. Additionally, we know that there is certainly 850\,$\mu$m flux associated with Bootes27 that is missed by the SCUBA-2 observations, as the fainter resolved SMA counterpart is significantly offset from the observed SCUBA-2 flux and has no apparent cross-match at 850\,$\mu$m. 

\subsection{FIR/Sub-mm Colours}
\label{sec: colours}

\begin{figure}
\centering
\includegraphics[width=0.45\textwidth]{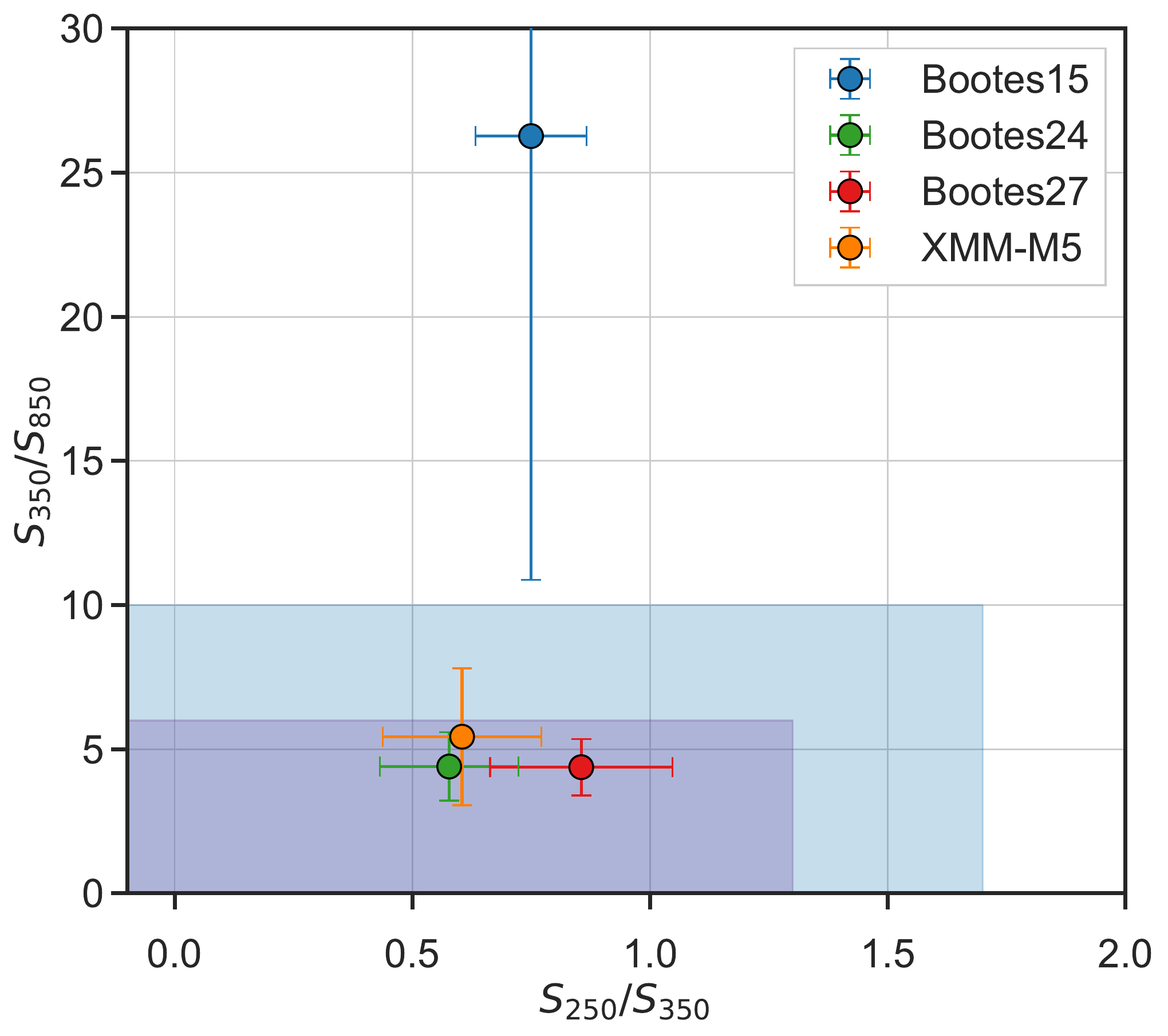}
\caption{$S_{250} / S_{350}$ vs. $S_{350} / S_{850}$ colour-colour diagram for our four 500\,$\mu$m risers. The purple region in the lower left corner, and the blue region surrounding it, indicate likely colours for $z \geq 2$ and $1 < z < 2$ DSFGs respectively. Bootes24, Bootes27 and XMM-M5 are all consistent with residing at $z \geq 2$, while Bootes15 remains unconstrained due to the poor SCUBA-2 photometry.}
\label{fig: 500_riser_colours}
\end{figure}

We have found that $3/4$ of our 500\,$\mu$m risers suffer from blending, which likely influences their FIR/sub-mm SEDs in numerous ways. Firstly, rather than being intrinsically red due to high redshifts ($z > 4$), it is likely that our 500\,$\mu$m risers are artificially reddened in the \textit{Herschel} SPIRE bands due to the fact that the successively larger beamsizes at longer wavelengths exacerbate the effects of blending \cite[see e.g.][]{2015ApJ...812...43B,2018MNRAS.477.1099D}. For example, \cite{2019ApJS..244...30M} and \cite{2021MNRAS.505.5260M} estimate that $\sim 20\%$ and $\sim 25\%$ of their samples respectively would not pass the 500\,$\mu$m riser selection criterion after accounting for the effects of blending. We additionally find that our SCUBA-2 observations may miss a significant fraction of the integrated flux density at 850\,$\mu$m (particularly for Bootes15 and Bootes27) and that there may be additional faint sources contributing to the integrated FIR/sub-mm flux density that remain undetected in our SMA observations (particularly for Bootes27). The combination of missing flux at 850\,$\mu$m and stronger blending at 500\,$\mu$m would produce an artificially steep Rayleigh-Jeans tail, leading to significant difficulties in interpreting the resulting galaxy properties. Therefore, for a rigorous SED analysis, de-blending of the \textit{Herschel} SPIRE flux densities will be necessary. However, due to the dearth of multi-wavelength ancillary data in this region of the Bootes field and the uncertainty as to whether we have recovered all of the sub-mm sources associated with each region, accurate de-blending of the \textit{Herschel} SPIRE flux density would be extremely challenging and could potentially produce misleading results. Given these difficulties, we do not attempt a rigorous SED fitting in this paper, and instead we simply estimate the photometric redshifts of our 500\,$\mu$m risers based on their integrated $S_{250} / S_{350}$ vs. $S_{350} / S_{850}$ colours (Figure \ref{fig: 500_riser_colours}). The purple and blue areas highlight regions in the colour-colour plot where $z \geq 2$ and $1 < z < 2$ DSFGs likely lie based on photometry from typical DSFG templates measured at various redshifts (see \cite{2019MNRAS.490.3840C} for more details). 

As an independent estimate of the photometric redshift, we additionally run the {\sc MMpz} algorithm\footnote{\url{http://www.as.utexas.edu/~cmcasey/mmpz.html}} \citep{2020ApJ...900...68C} which finds the most likely redshift at which a galaxy resides by determining where its FIR/mm SED is most consistent with the observed $L_{\text{IR}} - \lambda_{\text{peak}}$ relation, where $\lambda_{\text{peak}}$ is the rest-frame wavelength at which the FIR/mm SED peaks \citep{2013ApJ...778..131L,2016ApJ...822...80S,2018ApJ...862...77C,2020ApJ...900...68C,2021arXiv211006930C}. We use the integrated FIR photometry for each source, including the \textit{Herschel} SPIRE and SCUBA-2 photometry, and obtain a photometric redshift PDF for each of our 500\,$\mu$m risers. The full PDF for each source can be found in Appendix \ref{sec: pdfs}. Based on the upper and lower 68\% credible intervals, we find likely photometric redshifts of $1.83^{+1.33}_{-1.80}$ for Bootes15, $3.03^{+1.02}_{-0.90}$ for Bootes24, $2.53^{+0.83}_{-0.70}$ for Bootes27 and $2.88^{+1.02}_{-0.90}$ for XMM-M5. These values are in good agreement with the FIR/sub-mm colours, and we infer that Bootes24, Bootes27 and XMM-M5 likely lie at $z \geq 2$. We note that the photometric redshift of Bootes15 remains essentially unconstrained due to the poor SCUBA-2 photometry.

\begin{figure*}
\centering
\includegraphics[width=\textwidth]{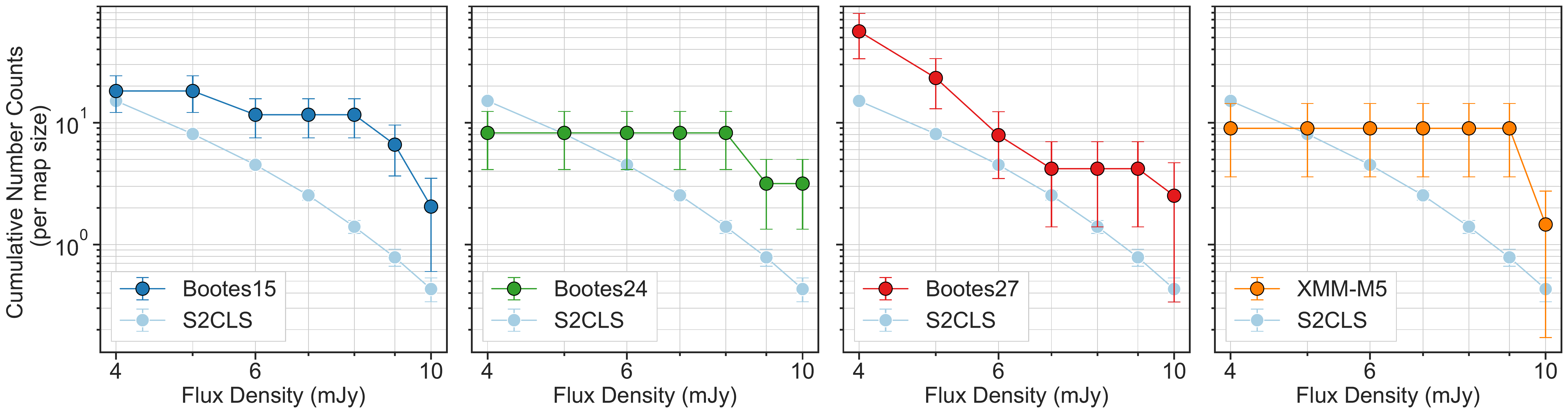}
\caption{Cumulative number counts for the four 500\,$\mu$m riser fields, plotted alongside the expected cumulative number counts in the field from S2CLS. The Bootes15, Bootes24 and XMM-M5 fields are overdense in bright ($S_{850} \gtrsim 8$\,mJy) sub-mm sources.}
\label{fig: number_counts}
\end{figure*}

\subsection{Wider Environment}

\subsubsection{Overdensities of 850\,$\mu$m Sources}
\label{sec: number_counts}

We then investigate the wider environment surrounding our four 500\,$\mu$m risers using the SCUBA-2 850\,$\mu$m maps reduced using the \texttt{REDUCE\_SCAN\_FAINT\_POINT\_SOURCES} recipe (as this reduction provides the highest signal-to-noise). We first crop the SCUBA-2 maps to a circle of diameter 700$^{\prime\prime}$ centred on the position of the SCUBA-2 detection, removing the edges of the map where the variance becomes much larger. We take each SNR map and find any pixel with $\text{SNR} > 4$, with connected pixels considered part of the same source. We then manually examine the extracted sources to ensure that only $\text{SNR} > 4$ detections are included and that there is no evidence of shredding (i.e. a single source being detected as multiple sources in the maps). In Appendix \ref{sec: scuba_maps} we present the reduced SCUBA-2 850\,$\mu$m maps and the positions of the $\text{SNR} > 4$ detected sources. For each source we extract deboosted flux densities and errors following the method outlined in Section \ref{sec: extracting_flux}, and construct cumulative number counts for our maps in steps of 1\,mJy. We note that the 500\,$\mu$m risers themselves are excluded from these cumulative number count calculations as they were selected to lie within the maps. We apply an effective area correction to account for the variable noise in the SCUBA-2 maps. For each of our detected SCUBA-2 sources, we take the extracted flux density prior to deboosting and find the threshold RMS noise value above which the source could no longer be detected with $\text{SNR} > 4$. We then calculate the number of pixels in the map with a standard deviation below this threshold value and multiply by the area of one pixel to calculate the effective area ($A_{e}$) in square arcseconds over which the source could be detected to SNR $ > 4$. We convert this effective area into units of `map size', where one map size is equal to the area of the 700$^{\prime\prime}$ diameter circular region. Each source then effectively contributes $1/A_{e}$ to the cumulative number counts \citep{2017ApJ...850...37W}. 

In order to determine whether there are overdensities of SMGs in our SCUBA-2 maps, we follow the analysis of \cite{2019MNRAS.490.3840C} and compare our cumulative number counts to the expected number of field counts from S2CLS. We first convert the cumulative number counts per square degree from Table 4 in \cite{2017MNRAS.465.1789G} to cumulative number counts per map size by multiplying by our map area in square degrees. In Figure \ref{fig: number_counts} we compare the cumulative number counts from S2CLS with our SCUBA-2 maps. We then quantify the overdensity at each flux density level using Poisson statistics. We first take the observed cumulative number counts above each flux density level and calculate the probability of observing that number of sources using the Poisson probability mass function

\begin{eqnarray}
    f(k) = \exp(-\mu)\frac{\mu^{k}}{k!},
\end{eqnarray}

where $k$ represents the number of observed counts and $\mu$ represents the expected number of counts from S2CLS. We then calculate the overdensity level using the equation

\begin{eqnarray}
    \delta = \frac{k - \mu}{\sqrt{\mu}},
\end{eqnarray}

\begin{table*}
    \centering
    \hspace{-0.1cm}
    \begin{tabular}{ccccccc}
    \hline\hline \\ [-2.0ex]
        Source & RA  & Dec & $S_{250\mu m}$ & $S_{350\mu m}$ & $S_{500\mu m}$ & $S_{850\mu m}$ \\ 
        & [J2000] & [J2000] & [mJy] & [mJy] & [mJy] & [mJy] \\
        \\ [-2.0ex]
        \hline\hline \\ [-2.0ex]
        Bootes15.SCUBA1 & 14:39:46.10 & $+$34:35:18.74
        & $0.0 \pm 3.5$ & $15.0 \pm 2.3$ & $13.4 \pm 2.7$ & $11.7 \pm 3.1$\\ [+1.0ex]
        Bootes15.SCUBA2 & 14:40:17.23 & $+$34:36:45.19
        & $6.8 \pm 2.8$ & $24.1 \pm 3.0$ & $0.0 \pm 4.4$ & $6.0 \pm 1.7$\\ [+1.0ex]
        Bootes15.SCUBA3 & 14:39:54.99 & $+$34:39:24.67
        & $25.0 \pm 2.3$ & $40.1 \pm 2.3$ & $32.9 \pm 3.0$ & $9.1 \pm 2.1$\\ [+1.0ex] \hline \\ [-2.0ex]
        Bootes24.SCUBA1 & 14:36:27.67 & $+$33:03:05.02
        & $46.1 \pm 2.2$ & $48.2 \pm 2.6$ & $28.7 \pm 4.1$ & $15.3 \pm 2.5$\\ [+1.0ex]
        Bootes24.SCUBA2 & 14:36:20.49 & $+$33:06:29.46
        & $73.7 \pm 2.3$ & $79.5 \pm 2.4$ & $57.0 \pm 3.2$ & $18.7 \pm 4.8$\\ [+1.0ex] \hline \\ [-2.0ex]
        Bootes27.SCUBA1 & 14:38:51.21 & $+$33:20:55.74
        & $19.6 \pm 2.3$ & $26.8 \pm 2.3$ & $29.0 \pm 2.7$ & $5.7 \pm 1.4$\\ [+1.0ex]
        Bootes27.SCUBA2 & 14:38:37.17 & $+$33:23:12.26
        & $47.9 \pm 2.3$ & $25.4 \pm 2.3$ & $26.5 \pm 3.2$ & $6.5 \pm 1.6$\\ [+1.0ex]
        Bootes27.SCUBA3 & 14:38:39.71 & $+$33:23:27.66
        & $564.2 \pm 4.6$ & $219.7 \pm 10.2$ & $83.8 \pm 17.8$ & $11.3 \pm 1.6$\\ [+1.0ex] \hline \\ [-2.0ex]
        XMM-M5.SCUBA1 & 02:19:14.18 & $-$04:37:38.09
        & $25.7 \pm 1.0$ & $37.7 \pm 1.1$ & $37.9 \pm 1.4$ & $10.5 \pm 3.5$\\ [+1.0ex]
        XMM-M5.SCUBA2 & 02:18:35.31 & $-$04:36:02.25
        & $15.4 \pm 1.1$ & $14.6 \pm 1.0$ & $7.7 \pm 1.3$ & $13.7 \pm 3.9$ \\ [+1.0ex]
        XMM-M5.SCUBA3 & 02:18:57.87 & $-$04:33:51.44
        & $81.5 \pm 1.1$ & $70.7 \pm 1.2$ & $40.9 \pm 1.9$ & $10.0 \pm 3.4$ \\ \\ [-2.0ex]
        \hline\hline 
    \end{tabular}
    \caption{Positions and integrated FIR/sub-mm photometry for the $\text{SNR} > 4$ SCUBA-2 detections with HerMES counterparts. \textit{Herschel} SPIRE flux densities are taken from the corresponding HerMES catalogue.}
    \label{tab: scuba_integrated_fir}
\end{table*}

as well as calculating an uncertainty in this overdensity by propagating the errors on the observed and expected counts.

We test the reliability in our SCUBA-2 maps using two separate methods. For the first method, we invert our SCUBA-2 point source reduced maps and extract the $>4\sigma$ negative noise peaks using the same method as outlined above. Assuming that there are roughly equal numbers of positive and negative spurious peaks in our SCUBA-2 maps, taking the ratio of these negative peaks to the total number of extracted positive sources should give us a reasonable estimate of how many of our sources are spurious. For the Bootes15 and XMM-M5 fields, the reliability is $\gtrsim 80\%$ at $4\sigma$, similar to other analyses of this nature \citep{2017MNRAS.468.4006M, 2019MNRAS.490.3840C}. For the Bootes24 and Bootes27 fields, the reliability drops to $\sim 60\%$ and $\sim 50\%$ respectively. The second method makes use of `jackknife' maps to estimate the reliability. First, we used the \texttt{MAKEMAP} command in the \texttt{SMURF} package to create an individual map for each of the available SCUBA-2 scans. We then invert half of these maps before co-adding all scans together to produce a single map with the sources removed (i.e. containing just noise). As with the previous reductions, we then applied a matched filter and subsequently produced a signal-to-noise map for each field. We find that the Bootes15 and Bootes24 fields both have a 100\% reliability at 4\,$\sigma$ based on the jackknife maps, while the Bootes27 field has a reliability of $\sim 88\%$. XMM-M5 has a somewhat lower reliability of 60\% using this method. The reliability estimates based on the jackknife maps are likely more robust as they do not rely on the assumption that there are the same number of spurious positive and negative noise spikes in the maps. We therefore include the reliability estimates from the jackknife maps in the uncertainties associated with the overdensity estimates.  We additionally tested the reliability for numerous different reductions in which we co-added various combinations of the available SCUBA-2 integrations for each field. The aforementioned reliability values, and the following discussions, are based on those combinations of integrations with the highest reliabilities.

To test the completeness of our SCUBA-2 maps we inserted fake sources at random positions within the SCUBA-2 map, re-ran the source extraction method and observed how many of the fake sources were recovered. The details of this completeness calculation are discussed in Appendix \ref{sec: completeness}. We find that Bootes15 and Bootes27 both reach a completeness of $\sim 50\%$ at $\sim 10$\,mJy, and rise to a completeness of $\gtrsim 80\%$ at 14\,mJy. By comparison, Bootes24 and XMM-M5 have a slightly poorer completeness, reaching $\sim 80$\% at $\sim 17$\,mJy and $\sim 16$\,mJy respectively. 
We find that Bootes15 hosts an $8.6 \pm 3.5$\,$\sigma$ overdensity of SMGs brighter than 8\,mJy, while Bootes24 shows a more marginal $5.8 \pm 3.5$\,$\sigma$ overdensity of SMGs brighter than 8\,mJy. The probabilities of observing these numbers of sources in a random field are $2.8 \times 10^{-7}$ and $1.1 \times 10^{-4}$ respectively. While Bootes27 appears to harbour a large overdensity of faint ($S_{850} > 4$\,mJy) sources, this overdensity is driven by a single source with a deboosted flux density of $\sim 4.9$\,mJy that would only be detected to $\text{SNR} > 4$ in $\sim 3\%$ of the total map area. This results in a large effective area correction and hence an artificially high overdensity value. If this single source is removed, then we do not observe any overdensity in the Bootes27 region. For XMM-M5, we find an overdensity of $9.3 \pm 6.5$\,$\sigma$ of sources brighter than 9\,mJy compared to the field counts, with a probability of observing this number of counts in a blank field of $\sim 2 \times 10^{-6}$. This analysis therefore identifies Bootes15, Bootes24 and XMM-M5 as regions plausibly containing overdensities of bright SCUBA-2 850\,$\mu$m sources.

\subsubsection{Cross-matching with HerMES Sources}

\begin{figure*}
\centering
\includegraphics[width=0.99\textwidth]{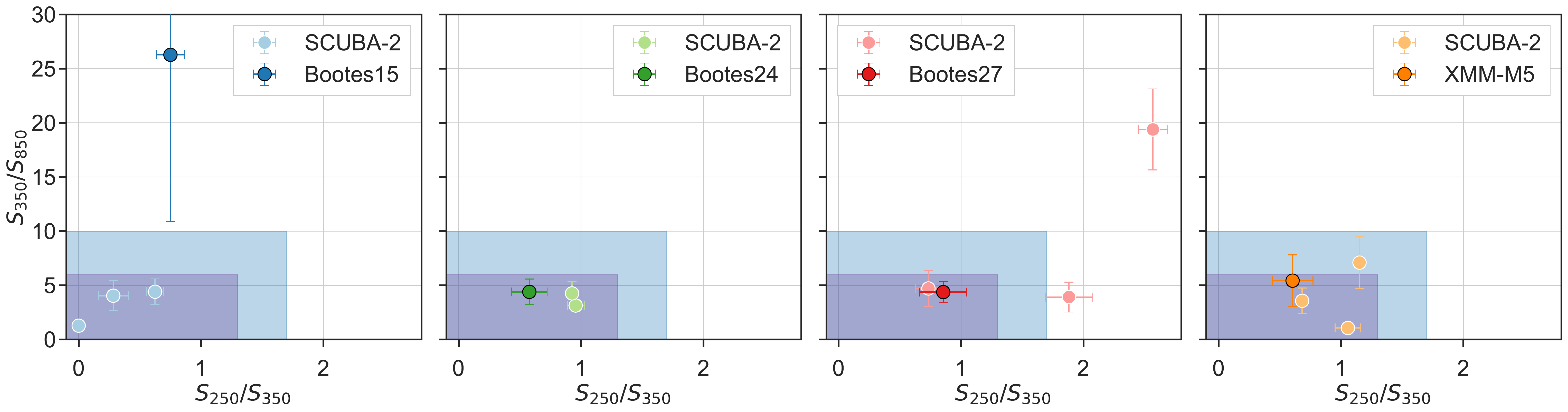}
\caption{The $S_{250} / S_{350}$ vs. $S_{350} / S_{850}$ FIR/sub-mm colours for our four 500\,$\mu$m risers (dark points) alongside the $\text{SNR} > 4$ detections in the SCUBA-2 850\,$\mu$m maps with \textit{Herschel} counterparts (light points). The purple region in the lower left corner, and the blue region surrounding it, indicate likely colours for $z \geq 2$ and $1 < z < 2$ DSFGs respectively. These plots demonstrate that the overdensities associated with Bootes24 and XMM-M5 are potentially physically associated at $z \geq 2$.}
\label{fig: colours}
\end{figure*}

\begin{figure*}
\centering
\includegraphics[width=0.99\textwidth]{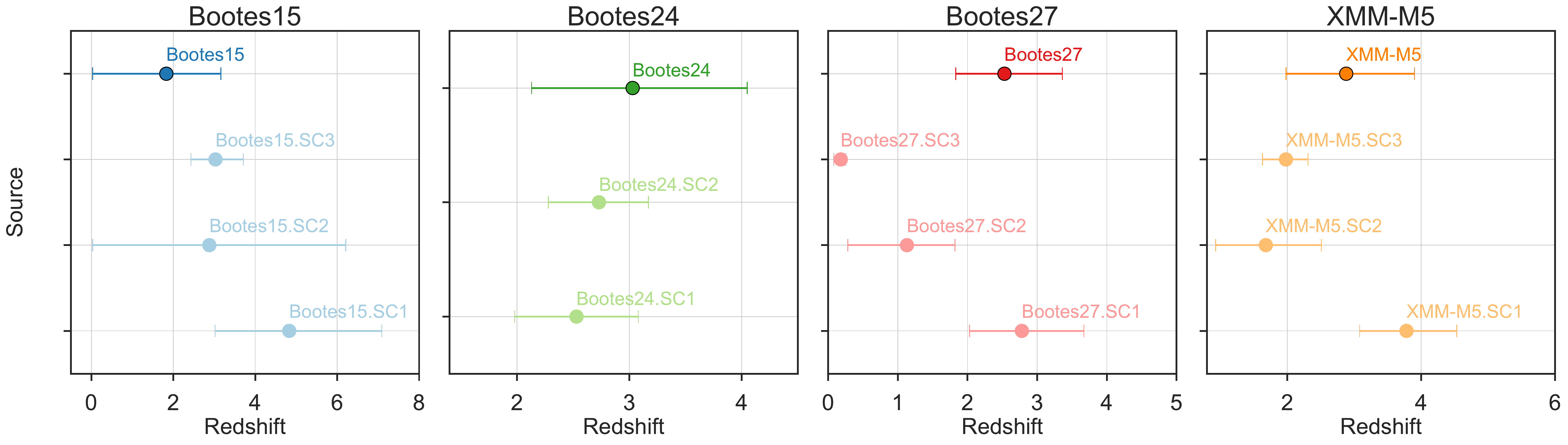}
\caption{The photometric redshifts for our 500\,$\mu$m risers and SCUBA-2 $\text{SNR} > 4$ detected sources estimated by {\sc MMpz}. Similarly to the FIR/sub-mm colours, we find that the Bootes24 region is consistent with containing a physically associated overdensity of bright DSFGs, with well constrained photometric redshifts placing the surrounding SMGs at $z \approx 2 - 3$, and Bootes24 itself at $z \approx 2 - 4$. While the photometric redshift of XMM-M5 does overlap with those of the bright SCUBA-2 sources, there is a larger range of photometric redshifts, indicating that this region may be somewhat more complex. The DSFGs associated with Bootes15 do not have well constrained photometric redshifts, and the bright SCUBA-2 sources in the Bootes27 region are likely unassociated with each other.}
\label{fig: mmpz_redshifts}
\end{figure*}

To investigate these potential overdensities further, we then cross-match the SCUBA-2 850\,$\mu$m detected sources with the HerMES catalogues. We use the optimum search radius of 9$^{\prime\prime}$ identified by \cite{2019MNRAS.490.3840C},  matching half of the FWHM of the \textit{Herschel} SPIRE 250\,$\mu$m beam, such that any SCUBA-2 source within 9$^{\prime\prime}$ of a source in the HerMES catalogue is considered a potential companion. Through this cross-matching process we find that $3/9$ (33\%), $2/4$ (50\%), $3/7$ (43\%) and  $3/5$ (60\%) SCUBA-2 detections have \textit{Herschel} counterparts for the Bootes15, Bootes24, Bootes27 and XMM-M5 fields respectively. These numbers exclude the 500\,$\mu$m risers themselves. The positions and FIR/sub-mm photometry for each of these sources can be found in Table \ref{tab: scuba_integrated_fir}. In Figure \ref{fig: colours} we present the $S_{250} / S_{350}$ vs. $S_{350} / S_{850}$ colours for both the 500\,$\mu$m risers and the SCUBA-2 detections, while in Figure \ref{fig: mmpz_redshifts} we plot their photometric redshifts and associated errors as estimated by {\sc MMpz} based on their \textit{Herschel} and SCUBA-2 photometry. The full photometric redshift PDFs from {\sc MMpz} can be found in Appendix \ref{sec: pdfs}.

The photometric redshift of Bootes15 is essentially unconstrained due to its poor SCUBA-2 photometry. However, the three SCUBA-2 detected sources with HerMES counterparts in the Bootes15 field have FIR/sub-mm colours consistent with $z \geq 2$, although {\sc MMpz} has trouble constraining the redshifts of two of them. Both Bootes24 and its companion SCUBA-2 detections have colours consistent with $z \geq 2$, and {\sc MMpz} places these sources at $z \sim 2-3$ (although Bootes24 itself suffers from larger error bars, placing it between $z \sim 2-4$). The FIR/sub-mm colours of sources in the Bootes27 region suggest that they are spread over a large range in redshift and so are unlikely to be physically associated, and this is supported by the results of {\sc MMpz}. The XMM-M5 field is plausibly overdense in FIR/sub-mm bright sources, and their colours suggest that they all reside at $z \geq 2$. The {\sc MMpz} results, however, offer a somewhat more complicated picture, placing two of the SCUBA-2 detections at $z \sim 1.5 - 2$ and the third (XMM-M5.SCUBA1) at $z \sim 4$. This is consistent with the \textit{Herschel} SPIRE photometry for XMM-M5.SCUBA1 which satisfies the colour criteria for a 500\,$\mu$m riser (see Table \ref{tab: scuba_integrated_fir}). 

Finally, we note that, for Bootes27,  one of the $\text{SNR} > 4$ SCUBA detections lies within $\sim 31^{\prime\prime}$ of the Bootes27 500\,$\mu$m riser, placing it just within the FWHM of the \textit{Herschel} SPIRE beamsize at 500\,$\mu$m ($35.2^{\prime\prime}$) but outside of the 250\,$\mu$m ($17.6^{\prime\prime}$) and 350\,$\mu$m ($23.9^{\prime\prime}$) beamsize. This SCUBA-2 source does not have a separate counterpart in the HerMES catalogues, indicating that it may contribute to making Bootes27 artificially red. For the remaining 500\,$\mu$m risers, the closest $\text{SNR} > 4$ detected SCUBA-2 source is too distant to be blended within the \textit{Herschel} SPIRE beamsize.  

\section{Discussion}
\label{sec: discussion}

\subsection{Multiplicities}
\label{sec: multiplicities}

We find evidence of multiplicity in three of our four 500\,$\mu$m risers, with Bootes15, Bootes24 and Bootes27 each breaking up into two resolved sources in the high resolution SMA maps, and all three showing extended emission in the SCUBA-2 maps. This builds on the work of G20, who found that $\sim 35\%$ of bright ($S_{500} > 60$\,mJy) 500\,$\mu$m risers are multiples, while fainter 500\,$\mu$m risers are also blends more often than expected. These conclusions were reached based on the assumption that the twelve 500\,$\mu$m risers without SMA detections in their follow-up observations are comprised of multiple DSFGs that were individually too faint to be detected in their SMA maps, but all of which contribute to the bright \textit{Herschel} source. By identifying evidence of multiplicity in three out of four of these 500\,$\mu$m risers, our observations support this conclusion. It is worth noting that, while we do observe evidence of multiplicity, we currently do not detect the high multiplicities of $\geq 3$ that were predicted by G20 for these 500\,$\mu$m risers. However, given the discrepancies between the \textit{Herschel}, SCUBA-2 and SMA photometry, we cannot rule out the possibility of additional faint counterparts lying below our current detection limits. In the following sections, we discuss Bootes15 and Bootes27, which are particularly interesting cases, in more detail. 

\subsubsection{Bootes15}
\label{sec: bootes15_multiplicity}

Figure \ref{fig: interferometer_vs_single-dish} demonstrates that the SMA and SCUBA-2 photometry for Bootes15 are in good agreement within the rather large 850\,$\mu$m error bars, but Figure \ref{fig: 850v500} indicates that its 850\,$\mu$m flux density is much lower than we would expect based on its 500\,$\mu$m flux density. The SCUBA-2 observations reach a depth of 2.2\,mJy, and so it is reasonable to expect that at least part of this discrepancy could be due to the SCUBA-2 observations missing a significant fraction of the total 850\,$\mu$m flux density in the system. This missing flux density could be in the form of extended, diffuse emission or additional point-like sources lying below our SCUBA-2 detection threshold. However, if we assume that the 850\,$\mu$m flux density is underestimated and in reality is more in line with similarly bright \textit{Herschel} sources in the S2CLS and STUDIES samples (e.g. $\gtrsim 10$\,mJy at 850\,$\mu$m), this would introduce a significant discrepancy between the SMA and SCUBA-2 photometry, similar to the discrepancy we observe in Bootes27. We therefore suggest that this discrepancy is caused not only by missing 850\,$\mu$m flux density in the SCUBA-2 observations, but also in part by faint sources lying below both our SMA and SCUBA-2 detection limits. Additional undetected sources would not only account for the faint SMA and SCUBA-2 flux densities, but the effects of blending would also artificially redden the \textit{Herschel} SPIRE photometry. \cite{2015ApJ...812...43B} find that blending can have a significant effect on the FIR/sub-mm colours of bright \textit{Herschel} sources. For example, blends of three sources can boost the $S_{500} / S_{350}$ colours of bright \textit{Herschel} sources by as much as $\sim 1.8$ times the deblended colours. This interpretation can therefore simultaneously account for the bright \textit{Herschel} SPIRE source associated with Bootes15, its red \textit{Herschel} colours and its unexpectedly faint SCUBA-2 and SMA photometry.

\subsubsection{Bootes27}
\label{sec: bootes27_multiplicity}

Bootes27 is another interesting case study. While the bright \textit{Herschel} 500\,$\mu$m riser does resolve into two individual SMA sources, one of these is much fainter than the other, remains undetected in the SCUBA-2 map and is offset from the detected SCUBA-2 850\,$\mu$m emission by $\sim 10^{\prime\prime}$. The SCUBA-2 contours show two partially resolved sources, a southern component coincident with the brighter SMA source, and a northern component with no apparent counterpart in the SMA images. Moreover, even excluding the partially resolved northern SCUBA-2 component, the brighter southern component has a peak, deboosted flux density of $\sim 6.9$\,mJy, significantly larger than the $1.9 \pm 0.3$\,mJy resolved SMA source associated with it. We also note that the single resolved SMA source is slightly offset from the brightest pixel in the SCUBA-2 maps. We interpret this tension between the SCUBA-2 and SMA observations as evidence of additional faint sources lying below even our deeper SMA detection threshold.

We can make a rough estimate of the number of sources that likely remain undetected in Bootes27 by comparing the integrated flux density from the SCUBA-2 observations to the resolved flux density from the SMA observations. The dirty map for Bootes27 that we used for source extraction has a $1\sigma$ RMS noise of $\sim 0.29$\,mJy/beam, meaning that any source with a flux density of less than $\sim 1.1$\,mJy would fall below the required $3.75\sigma$ detection threshold. The single resolved SMA source has a flux density of $1.9 \pm 0.3$\,mJy. Based on the integrated SCUBA-2 flux density of $10.1 \pm 1.6$\,mJy, we require at least 2.5\,mJy of additional flux density in our SMA maps in order to avoid a 3$\sigma$ tension between the SMA and SCUBA-2 photometry, corresponding to at least three additional faint sources lying below our SMA detection threshold. Alternatively, if we take the flux density estimates at face value, the SMA observations miss $8.2 \pm 1.6$\,mJy of flux density, and we would therefore require $\sim 6 - 9$ undetected SMA sources contributing to the total SCUBA-2 flux density. Such a high multiplicity would make Bootes27 a candidate high redshift protocluster core, potentially similar to those discovered by \cite{2018Natur.556..469M} and \cite{2018ApJ...856...72O}. We note that the ancillary \textit{Spitzer} IRAC data for Bootes27 shows $\sim 7$ individual sources detected to $>3\sigma$ at 8.0\,$\mu$m, $\sim 3$ of which remain completely undetected at 4.5\,$\mu$m, indicating that there may be additional, extremely red sources contributing to the SCUBA-2 flux density that are undetected in the SMA map. A more comprehensive, multi-wavelength analysis of Bootes27 will be presented in Cairns et al. (in prep).

\subsection{Photometric Redshifts}
\label{sec: photo-z}

Based on their integrated FIR/sub-mm colours in Figure \ref{fig: 500_riser_colours}, we find that our 500\,$\mu$m risers likely lie at $z \geq 2$, excluding Bootes15 whose FIR/sub-mm colours remain poorly constrained due to the uncertain 850\,$\mu$m photometry. This result is corroborated by {\sc MMpz}, which indicates that these 500\,$\mu$m risers likely lie at $z \sim 2 - 4$. These photometric redshifts are consistent with our interpretation of these sources as blends of multiple DSFGs. As well as being effective at selecting intrinsically red, high redshift DSFGs, the 500\,$\mu$m riser selection criterion will also naturally include some fraction of intermediate redshift systems that are artificially reddened by the blending of multiple sources at moderately large separations. For example, \cite{2021MNRAS.505.5260M} find in their sample of $\sim 100$ 500\,$\mu$m risers that the photometric redshift distribution of those that resolve into multiple components is skewed towards lower redshifts, with a slightly lower median redshift ($z _{\text{med}} = 3.5$ vs. $z _{\text{med}} = 3.8$) and a higher fraction residing at lower redshifts (27\% at $z_{\text{phot}} < 3$ vs. 10\%) compared to single systems. We find that our 500\,$\mu$m risers typically fall into the category of multiple systems at moderate redshifts and, based on their FIR/sub-mm flux densities and intermediate photometric redshifts, we would expect them to be examples of Hyper-Luminous Infrared Galaxies (HLIRGS: $L_{\text{FIR}} > 10^{13}$\,L$_{\odot}$), similar to those discussed in G20.

\subsection{Environments}
\label{sec: environments}

Our SCUBA-2 850\,$\mu$m observations map out sub-mm emission over a $>10'$ region surrounding our 500\,$\mu$m risers, and are therefore extremely useful for characterising their environments. DSFGs have the potential to be used as signposts for overdensities of galaxies in the early Universe, allowing us to identify the potential progenitors of massive, local galaxy clusters. We find that, while the Bootes27 region is consistent with the expected number of field counts from S2CLS, Bootes15, Bootes24 and XMM-M5 are all consistent with lying in overdensities of SMGs with 850\,$\mu$m flux densities greater than $\sim 8$\,mJy. We do note, however, that the error bars on these overdensity values for Bootes24 and XMM-M5 are rather large, with only a very marginal overdensity at the lower limit. This result is interesting as we would not necessarily expect our 500\,$\mu$m risers to be effective tracers of high-redshift galaxy overdensities - based on the clustering analysis of SMGs from the ALESS survey, \cite{2020ApJ...904....2G} find that only the brightest SMGs ($S_{870} > 5 - 6$\,mJy) should trace massive structures at $z \sim 2$, while our sources are significantly fainter than this in the resolved SMA maps. Without spectroscopic redshifts for the SCUBA-2 sources we cannot say with certainty whether these overdensities are physically associated structures or simply chance line of sight projections of bright sub-mm sources. In the absence of spectroscopic data, we must rely on photometric redshifts estimated from FIR/sub-mm colour-colour diagrams and the {\sc MMpz} algorithm. From Figures \ref{fig: colours} and \ref{fig: mmpz_redshifts}, we infer that the marginal overdensity associated with Bootes24 could be physically associated, with the FIR/sub-mm colours placing the sources at $z \geq 2$, and {\sc MMpz} placing them at $z \sim 2-4$. For the Bootes15 field, improvements in the FIR/sub-mm photometry are required to better constrain the photometric redshifts of the detected sources, while the sources in the Bootes27 field likely reside at very different redshifts. The SCUBA-2 detections in the XMM-M5 field could represent an overdensity of SMGs, and their colours suggest that they are all at $z \geq 2$, but {\sc MMpz} places two of these sources at $z \sim 1.5 - 2$, and the third at $z \sim 4$. This overdensity is therefore less likely to be a physically associated structure than Bootes24. Additional observations providing improved photometry and/or spectroscopic redshifts for these sources will be required to conclude with certainty whether they are physically associated structures.

\subsection{Comparison to Predictions for Bright \textit{Herschel} Sources}
\label{sec: simulations}

\subsubsection{Multiplicity Fraction}

Our results support the conclusions of G20, who find that $\sim 35\%$ of bright ($S_{500} > 60$\,mJy) 500\,$\mu$m risers, and $\sim 60\%$ of faint ($S_{500} < 60$\,mJy) 500\,$\mu$m risers are multiples, although we note that these observations were somewhat inhomogeneous, making use of a variety of frequencies (ranging from 231\,GHz to 346\,GHz) and configurations, with a mixture of full tracks, partial tracks and track sharing, resulting in a range of angular resolutions (with an average beam FWHM of $2.5 \pm 0.8$ arcseconds) and depths ($0.5-2.9$\,mJy/beam).

\cite{2017arXiv170904191O} provide high-resolution ($\sim 0.12^{\prime\prime}$) ALMA follow-up observations at 870\,$\mu$m for 44 ultra-red DSFGs from HerMES and H-ATLAS with $S_{500} / S_{250} > 1.5$ and $S_{500} / S_{350} > 1.0$, reaching $1\sigma$ RMS sensitivities of 0.1\,mJy/beam. They find that, in total, $\approx 39\%$ resolve into multiple components and, more specifically, $\approx 22\%$ of bright ($S_{500} > 60$\,mJy) and $\approx 65\%$ of faint ($S_{500} \leq 60$\,mJy) sources in their sample resolve into multiple components.

\cite{2017A&A...607A..89B} produce a FIR to sub-millimeter simulation of the extragalactic sky based on a galaxy evolution model with two star-formation modes. They extract sources from maps based on these observations to investigate the effects that the limited angular resolution of single-dish instruments has on observations in the FIR/sub-mm. They predict that all \textit{Herschel} sources brighter than 60\,mJy at 500\,$\mu$m should be individual, lensed sources, while $\sim40\%$ of sources fainter than 60\,mJy at 500\,$\mu$m should be blended. 

In their compilation of 63 500\,$\mu$m risers, \cite{2019ApJS..244...30M} estimate a multiplicity fraction of $\sim 27\%$ based on numerous follow-up observations with NOEMA, ALMA and the SMA, although these studies reach various depths and angular resolutions, as well as probing much fainter 500\,$\mu$m flux densities. This multiplicity fraction rises to $\sim 39\%$ when sources showing evidence of strong gravitational lensing are removed.

\begin{figure*}
\includegraphics[width=\textwidth]{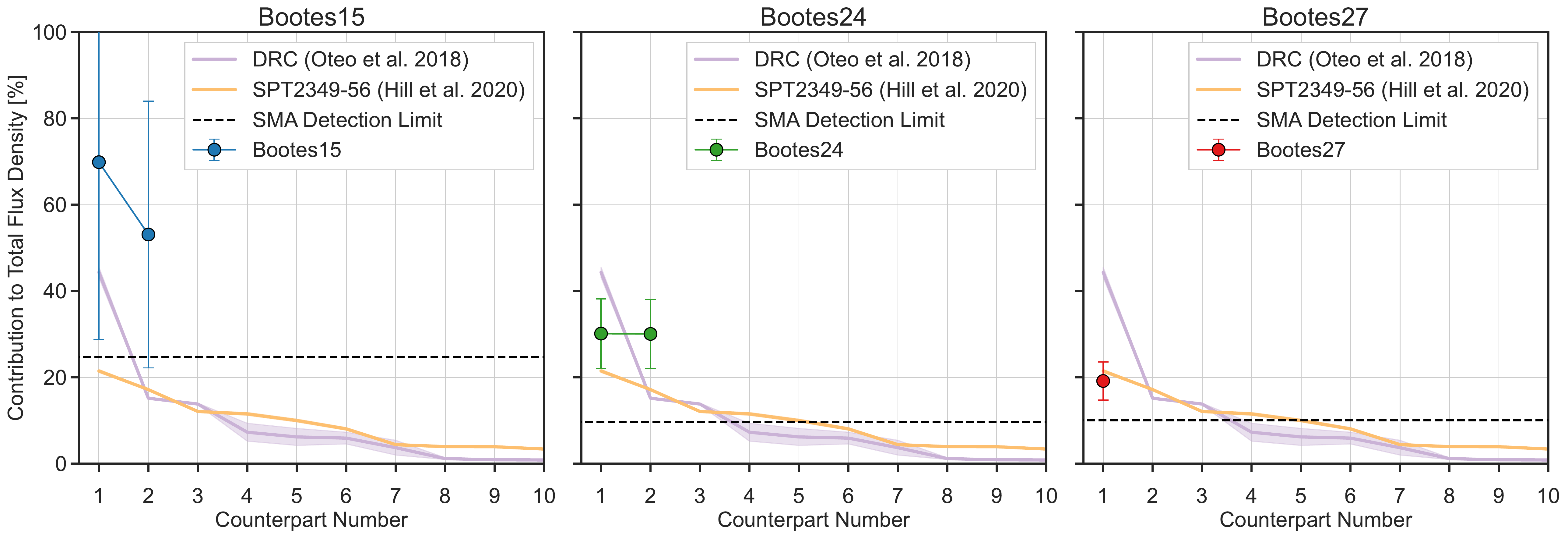}
\caption{The relative contributions of each of our SMA detections to the total integrated SCUBA-2 850\,$\mu$m flux density. The orange and purple lines show the flux density distributions of the DRC and SPT2349-56 protoclusters respectively, while the black dashed lines show the $3.75\sigma$ detection limits in our dirty SMA maps.  The contribution of each SMA source to the integrated SCUBA-2 flux density of Bootes15 remains poorly constrained. The SMA sources associated with Bootes24 and Bootes27 recover $60 \pm 15\%$ and $19 \pm 4\%$ of the total integrated SCUBA-2 flux density and so there may be additional sources residing below our SMA detection limits. However, if further sources do reside below our detection threshold, they likely follow a different flux density distribution to the DRC and SPT2349-56 protoclusters.}
\label{fig: flux_distribution}
\end{figure*}

\cite{2021MNRAS.505.5260M} carry out 1.1\,mm LMT/AzTEC observations of $\sim 100$ 500\,$\mu$m risers with $S_{500} \approx 35 - 80$\,mJy, achieving an average beam FWHM of $9.6 \pm 0.5$ arcseconds and $1\sigma$ RMS sensitivities of $\approx 0.7-2.8$\,mJy. They find that $\sim 9\%$ show direct evidence of multiplicity, rising to $\sim23\%$ if 500\,$\mu$m risers which display evidence of multiplicity but are slightly below their detection threshold are included, and up to $\sim50\%$ if 500\,$\mu$m risers without counterparts in their higher resolution observations are assumed to be multiple sources. This is considered an extreme scenario as a higher dust spectral emissivity index $\beta$ could also render some of their sources undetectable at 1.1\,mm, but our observations of 500\,$\mu$m risers without counterparts in previous, high resolution follow-up observations indicate that a significant fraction of these non-detections could be multiple systems. For the twelve sources in their sample with $S_{500} > 60$\,mJy, two break up into multiple components and three show no counterparts in their higher resolution images, indicating a multiplicity of $\approx 17-42\%$. However, the larger beamsize of their observations compared to \cite{2017arXiv170904191O} and G20 probe multiplicity on a significantly different scale, such that smaller scale multiplicities would likely be classified as individual sources. They estimate a correction factor of $\sim 10\%$ accounting for multiple systems separated by distances much smaller than their $9.6 \pm 0.5$ arcsecond beamsize, bringing the multiplicity fraction up to  $\gtrsim 18\%$ in the conservative case and $\sim 60\%$ in the extreme case.

We find that our three Bootes sources all show some evidence of multiplicity, with all three resolving into two sources in the high-resolution SMA maps and showing an extended morphology in the SCUBA-2 maps. This result supports the conclusions of G20 and, by extension, is in good agreement with the multiplicity fractions for 500\,$\mu$m risers quoted by \cite{2017arXiv170904191O}, despite their ALMA observations achieving significantly better sensitivities and angular resolutions. These results are also in reasonably good agreement with the more extreme scenario in \cite{2021MNRAS.505.5260M}, and well within the range of potential multiplicity fractions of the brightest ($S_{500} > 60$\,mJy) sources in their sample. These multiplicity fractions indicate that 500\,$\mu$m risers are a much more diverse population than previously predicted. 

\subsubsection{Brightest-Galaxy Fraction}

We can also compare the flux distributions of the resolved SMA sources in our 500\,$\mu$m risers to the literature. \cite{2017A&A...607A..89B}, based on their simulations, predict that the brightest galaxy should contribute $\sim 60\%$ of the total 500\,$\mu$m flux density. Similarly, \cite{2018A&A...614A..33D} find that the brightest galaxy inside the \textit{Herschel} beam for their simulated SPIRE data contributes on average $75\%$ and $64\%$ at 250\,$\mu$m and 500\,$\mu$m respectively. \cite{2021MNRAS.505.5260M} find that the brightest component contributes on average $50 - 75\%$ of the total flux density at 1.1\,mm, while \cite{2019ApJS..244...30M} find that the brightest ALMA source contributes on average $41-80\%$ of the total ALMA flux density and $15-59\%$ of the total SCUBA-2/LABOCA flux density at a similar wavelength. 

We find that the brightest SMA component in Bootes15 contributes $56 \pm 12\%$ to the total SMA flux density, and $69 \pm 41\%$ of the integrated SCUBA-2 850\,$\mu$m flux density. However, as discussed in Section \ref{sec: SMA_SCUBA_detection}, it is likely that we do not recover the total integrated FIR/sub-mm flux density of Bootes15 in our observations. For Bootes24 and Bootes27, the brightest SMA component contributes $50 \pm 9\%$ and $53 \pm 18\%$ respectively to their total SMA flux densities, but just $30 \pm 8\%$ and $19 \pm 4\%$ respectively to their integrated SCUBA-2 flux densities at a similar wavelength. These values are in good agreement with \cite{2019ApJS..244...30M}, and indicate that for Bootes24 and Bootes27 we may not recover the total sub-mm flux density. 

\subsubsection{Comparison to Known Protocluster Cores}

We can also compare these values to the flux density distributions of known high-$z$ protocluster cores, such as the Distant Red Core \cite[DRC:][]{2018ApJ...856...72O} and SPT2349-56 \citep{2018Natur.556..469M}. For the DRC, we focus on the ALMA 2\,mm follow-up observations carried out by \cite{2018ApJ...856...72O}, reaching depths of $\sim 6$\,$\mu$Jy/beam with a synthesised beam FWHM of $1.68^{\prime\prime} \times 1.54^{\prime\prime}$. For SPT2349-56, we use the flux densities extracted from the ALMA 850\,$\mu$m continuum maps based on the [{\sc Cii}] map channels with no line emission \citep{2020MNRAS.495.3124H}. In Figure \ref{fig: flux_distribution}, we then plot the contribution of each of the ten brightest resolved sources associated with the protocluster cores to the total flux density in the aforementioned bands. For our 500\,$\mu$m risers, we plot the contribution of each of the resolved SMA sources to the integrated SCUBA-2 850\,$\mu$m flux density, and estimate the uncertainties in these contributions by adding in quadrature the errors in the resolved SMA and integrated SCUBA-2 flux densities. Note that we exclude XMM-M5 from this plot, as the SMA and SCUBA-2 observations were taken at very different wavelengths. We additionally exclude the fainter SMA source associated with Bootes27, as it is offset from the observed SCUBA-2 emission and so likely does not contribute to the observed 850\,$\mu$m flux density.

For Bootes15, the contribution of each SMA source to the total SCUBA-2 flux density is poorly constrained, primarily due to the large uncertainties in the 850\,$\mu$m photometry. In total, we estimate that the SMA observations recover $123 \pm 73\%$ of the SCUBA-2 observations. However, as previously discussed, it is likely that some fraction of the total SCUBA-2 flux density is missed by our observations. The two SMA sources associated with Bootes24 recover $60 \pm 15\%$ of the total SCUBA-2 flux density, shared equally between the two resolved sources.  Therefore, the combined contribution from the two brightest sources is in good agreement with the two brightest sources in the DRC, which similarly contribute $\sim 60\%$ to the total flux density of the system. However, the flux density is shared much more unequally between the two brightest sources in the DRC. If there were additional sub-mm sources contributing to Bootes24, we would expect to detect one extra source in our SMA observations if they were to follow a DRC-like flux density distribution. Alternatively, we would expect to detect the brightest $4-5$ sources if they were to follow a SPT2349-56-like flux density distribution. For Bootes27, the single resolved SMA source associated with the observed SCUBA-2 emission contributes just $19 \pm 4\%$ to the integrated 850\,$\mu$m flux density. This is similar to the $\sim 22\%$ contribution to the total system from the brightest source in SPT2349-56. However, if Bootes27 were to represent a protocluster core with a similar flux density distribution to SPT2349-56, we would similarly expect to detect the brightest $4-5$ components, whereas we only detect a single source associated with the 850\,$\mu$m emission. 

We infer that Bootes24 and Bootes27 may contain additional, faint sources lying below our SMA detection threshold that contribute to the total integrated SCUBA-2 flux density. However, it seems that this flux density must be distributed between its members differently to known high redshift protocluster cores, with each of the fainter members individually contributing significantly less to the total sub-mm flux density than expected.

\section{Conclusions}
\label{sec: conclusions}

We have presented \textit{Herschel}, SMA and SCUBA-2 observations of four 500\,$\mu$m risers from the HerMES survey. These sources were selected based on their red \textit{Herschel} SPIRE colours, but remained undetected in previous high-resolution follow-up observations conducted by G20. The previous results indicated that they were likely comprised of multiple DSFGs that individually were faint enough to remain undetected in their high resolution SMA observations, but each contributing to a large integrated flux density when smoothed over the \textit{Herschel} SPIRE beamsize. In our deeper FIR/sub-mm observations, each source is detected to $>3\sigma$ in the SCUBA-2 maps, and there is at least one $>3.75\sigma$ detection in each of the SMA maps. Our findings can be summarised as follows.

\begin{enumerate}
    \item We find evidence of multiplicity in three out of four 500\,$\mu$m risers. Bootes15, Bootes24 and Bootes27 each break up into two faint sources in the high-resolution SMA images, while XMM-M5 resolves into a single source. Bootes27 additionally displays a bright, partially resolved SCUBA-2 detection without any apparent SMA counterparts, indicating that there may be further faint DSFGs lying below the SMA detection threshold. This is in line with the results of G20 who found (by assuming that their 500\,$\mu$m risers without SMA counterparts are sources with high multiplicities) that $\sim35\%$ of bright ($S_{500} > 60$\,mJy) 500\,$\mu$m risers and $\sim 60\%$ of faint ($S_{500} < 60$\,mJy) 500\,$\mu$m risers are multiples. By providing deeper SMA follow-up observations of these 500\,$\mu$m risers, we confirm that they are likely comprised of multiple faint components. These results indicate that the 500\,$\mu$m riser population is significantly more diverse than expected.
    
    \item For the three 500\,$\mu$m risers in the Bootes field, the SMA observations recover $123 \pm 73\%$, $60 \pm 15\%$ and $19 \pm 4\%$ of the integrated SCUBA-2 flux density respectively indicating that, for Bootes24 and Bootes27, there may be additional, faint sources below the current SMA detection limit, but contributing to the integrated SCUBA-2 flux density. In particular, for Bootes27, there is a $>3\sigma$ disparity between the resolved SMA and integrated SCUBA-2 flux densities. We estimate that at least three additional faint sources are required to recover the total SCUBA-2 flux density. This could make Bootes24 and Bootes27 examples of dense, protocluster cores of DSFGs, only a handful of which have been discovered to date.
    
    \item It is likely that our SCUBA-2 observations miss some sub-mm flux density, particularly for Bootes15, either in the form of extended diffuse emission or additional, faint, point-like sources. For Bootes27, there is a resolved SMA source offset from the observed SCUBA-2 emission and without an apparent SCUBA-2 counterpart.
    
    \item We estimate the photometric redshifts of our 500\,$\mu$m risers using their FIR/sub-mm colours, as well as the {\sc MMpz} algorithm. While the photometric redshift of Bootes15 remains poorly constrained, Bootes24, Bootes27 and XMM-M5 all likely lie at $z \geq 2$. This is consistent with the interpretation that the 500\,$\mu$m riser selection criterion recover both intrinsically red sources at $z > 4$ and artificially red, multiple systems at more moderate redshifts. Based on their FIR flux densities and photometric redshifts, we expect our sources to be HLIRGs, similar to those in the G20 sample.

    \item By comparing the cumulative number counts of $\text{SNR} > 4$ detections in the SCUBA-2 850\,$\mu$m maps with the expected number of field counts from the SCUBA-2 Cosmology Legacy Survey, we find that Bootes15, Bootes24 and XMM-M5 lie in regions that are $8.6 \pm 3.5$\,$\sigma$, $5.8 \pm 3.5$\,$\sigma$ and $9.3 \pm 6.5$\,$\sigma$ overdense in bright ($\gtrsim 8$\,mJy) SMGs respectively. For those SCUBA-2 detections that have \textit{Herschel} SPIRE counterparts, we estimate photometric redshifts based on their FIR/sub-mm photometry. We find that the overdensity associated with Bootes24 could feasibly be physically associated, while Bootes15 and XMM-M5 are more likely to be chance line-of-sight arrangements of sources.
    
    \item We compare the inferred multiplicity fractions of 500\,$\mu$m risers from G20 to similar studies in the literature, finding that they are in good agreement with \cite{2017arXiv170904191O} and with the more extreme scenarios in \cite{2021MNRAS.505.5260M}. We also find that the brightest SMA component of Bootes24 and Bootes27 contributes a similar amount to the total integrated SCUBA-2 flux density as the brightest resolved source associated with the 500\,$\mu$m risers in the  \cite{2019ApJS..244...30M} sample.
    
    \item We also compare the flux density distributions of our resolved SMA sources to those of known protocluster cores. We find that, while the SMA sources associated with Bootes24 and Bootes27 only recover a fraction of the total sub-mm flux density, we would have expected to detect the brightest $4-5$ components if the multiple systems followed a similar flux density distribution to SPT2349-56, or the brightest three components if they followed a similar flux density distribution to the DRC. Therefore, if these 500\,$\mu$m risers are examples of dense protocluster cores, they likely follow a different flux density distribution to known, high-$z$ protocluster cores.
    
\end{enumerate}

We therefore find that our 500\,$\mu$m risers are typically comprised of multiple DSFGs and may represent examples of high redshift protocluster cores similar to those found by \cite{2018Natur.556..469M} and \cite{2018ApJ...856...72O}. However, additional follow-up observations will be required to confirm the nature of these sources, and there are a number of available avenues that these observations could take. Deeper sub-mm continuum/spectroscopic observations with NOEMA will likely be required to determine if there are any additional, faint, point-like sources below our current SMA detection limits that contribute to the integrated SCUBA-2 flux (and if these sources are physically associated), particularly for Bootes24 and Bootes27 where our SMA observations currently only recover a fraction of the total 850\,$\mu$m flux density. Alternatively, deep optical/NIR follow-up observations would allow us to accurately deblend the \textit{Herschel} SPIRE flux density, from which we could attempt a rigorous SED fitting for each of the resolved components and estimate galaxy properties such as SFR and stellar mass. Finally, additional spectroscopic follow-up observations for sources in the wider field would allow us to conclude whether any larger scale overdensities in these regions represent physically associated structures.

\section*{Acknowledgements}

The authors wish to recognize and acknowledge the very significant cultural role and reverence that the summit of Mauna Kea has always had within the indigenous Hawaiian community.  We are most fortunate to have the opportunity to conduct observations from this mountain.

The Submillimeter Array is a joint project between the Smithsonian Astrophysical Observatory and the Academia Sinica Institute of Astronomy and Astrophysics and is funded by the Smithsonian Institution and the Academia Sinica. 

The James Clerk Maxwell Telescope is operated by the East Asian Observatory on behalf of The National Astronomical Observatory of Japan; Academia Sinica Institute of Astronomy and Astrophysics; the Korea Astronomy and Space Science Institute; Center for Astronomical Mega-Science (as well as the National Key R\&D Program of China with No. 2017YFA0402700). Additional funding support is provided by the Science and Technology Facilities Council of the United Kingdom and participating universities and organizations in the United Kingdom and Canada. 

This research has made use of data from HerMES project (\url{http://hermes.sussex.ac.uk/}). HerMES is a Herschel Key Programme utilising Guaranteed Time from the SPIRE instrument team, ESAC scientists and a mission scientist.The HerMES data was accessed through the Herschel Database in Marseille (HeDaM - \url{http://hedam.lam.fr}) operated by CeSAM and hosted by the Laboratoire d'Astrophysique de Marseille.

This research used the facilities of the Canadian Astronomy Data Centre operated by the National Research Council of Canada with the support of the Canadian Space Agency.

This research made use of APLpy, an open-source plotting package for Python hosted at \url{http://aplpy.github.com}. This research made use of Astropy, a community-developed core Python package for Astronomy \citep{2013A&A...558A..33A,2018AJ....156..123A}. This research made use of Photutils, an Astropy package for
detection and photometry of astronomical sources \citep{larry_bradley_2021_5796924}. This research made use of the Starlink Table/VOTable Processing Software \texttt{TOPCAT} \citep{2005ASPC..347...29T}.

J.C. acknowledges support from the Science and Technology Facilities Council [grant number ST/S505432/1]. I.P.-F. acknowledges support from the Spanish State Research Agency (AEI) under grant number PID2019-105552RB- C43. 

\section*{Data Availability}
The data underlying this article will be shared on reasonable request to the corresponding author.

%%%%%%%%%%%%%%%%%%%%%%%%%%%%%%%%%%%%%%%%%%%%%%%%%%


\balance
\bibliographystyle{mnras}
\bibliography{references}

%%%%%%%%%%%%%%%%%%%%%%%%%%%%%%%%%%%%%%%%%%%%%%%%%%

%%%%%%%%%%%%%%%%% APPENDICES %%%%%%%%%%%%%%%%%%%%%
\appendix
\onecolumn

\section{Probability Density Functions}
\label{sec: pdfs}

\begin{figure*}
\centering
\includegraphics[width=\textwidth]{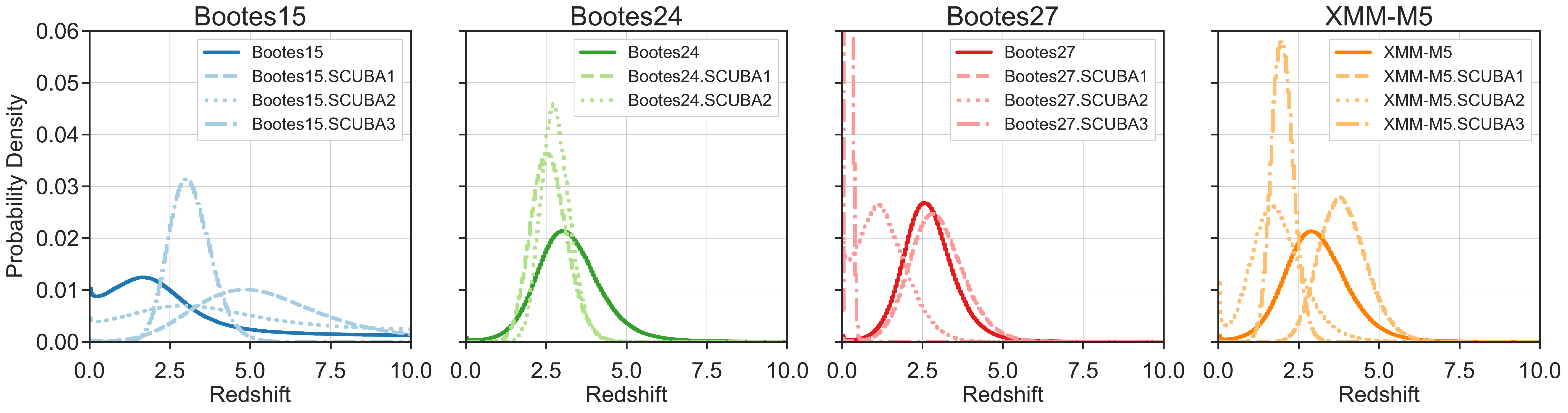}
\caption{Normalised photometric redshift PDFs for our 500\,$\mu$m risers and their associated $\text{SNR} > 4$ SCUBA-2 detections estimated by {\sc MMpz}.}
\label{fig: pdfs}
\end{figure*}

In Figure \ref{fig: pdfs}, we present the photometric redshift PDFs obtained from {\sc MMpz} for our 500\,$\mu$m risers and the other $\text{SNR} > 4$ detections in the SCUBA-2 fields. The {\sc MMpz} fitting algorithm finds the most likely redshift at which a galaxy resides by determining where its FIR/mm SED is most consistent with the observed $L_{\text{IR}} - \lambda_{\text{peak}}$ relation, where $\lambda_{\text{peak}}$ is the rest-frame wavelength at which the FIR/mm SED peaks \citep{2013ApJ...778..131L,2016ApJ...822...80S,2018ApJ...862...77C,2020ApJ...900...68C,2021arXiv211006930C}. We find that the photometric redshifts for Bootes15, Bootes15.SCUBA1 and Bootes15.SCUBA2 are poorly constrained due to poor quality photometry. By comparison, we find that the photometric redshift PDFs for the sources in the Bootes24 field are all well constrained, likely residing at redshifts of $z \sim 2 - 3$, with Bootes24 itself being somewhat broader with a higher redshift tail to its PDF. For Bootes27 and XMM-M5, the photometric redshift PDFs are similarly well constrained, but the sources are spread over a wider range of redshifts, with two of the SCUBA-2 sources likely residing at $z \sim 2$, one likely residing at $z \sim 4$, and XMM-M5 itself straddling these two redshifts. For Bootes24, Bootes27 and XMM-M5, these results are in good agreement with the FIR/sub-mm colours shown in Figure \ref{fig: colours}.

\section{SCUBA-2 Maps}
\label{sec: scuba_maps}

\begin{figure*}
\centering
\includegraphics[width=0.9\textwidth]{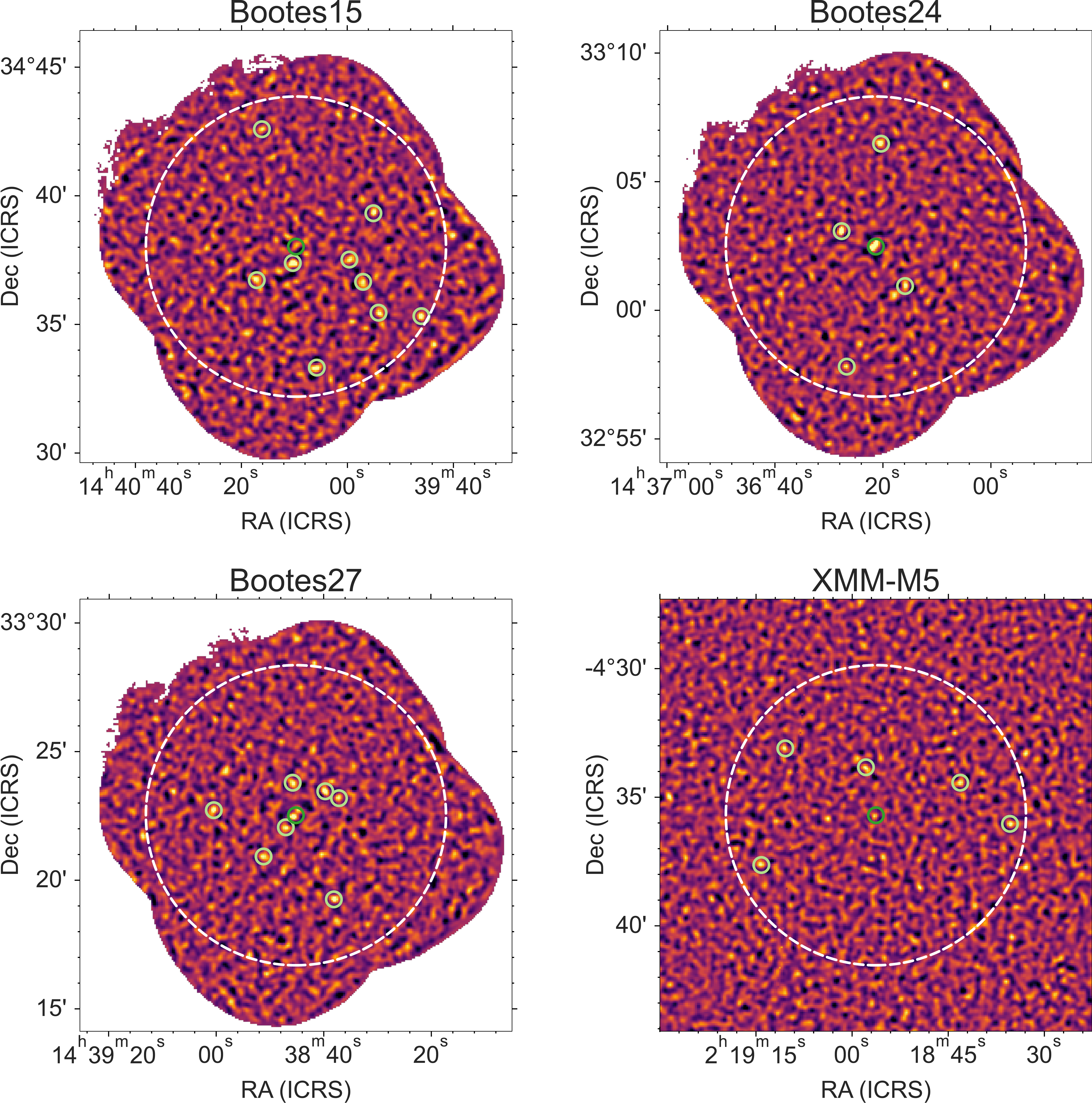}
\caption{SCUBA-2 850\,$\mu$m maps centred on each of our 500\,$\mu$m risers reduced using the \texttt{REDUCE\_SCAN\_FAINT\_POINT\_SOURCES} recipe. The darker circles show the positions of the 500\,$\mu$m risers, while the lighter circles show the positions of the $\text{SNR} > 4$ detected SCUBA-2 sources. The white dashed line shows the circle of radius $350^{\prime\prime}$ over which sources are extracted.}
\label{fig: scuba_maps}
\end{figure*}

In Figure \ref{fig: scuba_maps} we present the full SCUBA-2 850\,$\mu$m maps for the fields covering our four 500\,$\mu$m risers. These maps were reduced using the \texttt{REDUCE\_SCAN\_FAINT\_POINT\_SOURCES} recipe, and sources were extracted within the white dashed circle. We note that the SMGs are distributed differently within the different fields. Sources in Bootes15 are mostly clustered towards the South Western region of the map, while sources in Bootes27 seem to be more clustered around the centre. Sources in Bootes24 and XMM-M5 are more spread out across the map.

\section{Completeness}
\label{sec: completeness}

In this section, we discuss the details of our completeness calculation. For each of our SCUBA-2 maps, we inserted fake sources within the 700$^{\prime\prime}$ diameter circular region, where each fake source was defined as a simple 2D Gaussian with a random position, a defined peak flux density and a width corresponding to $\sigma = 7.3^{\prime\prime}$ (half of the FWHM of the SCUBA-2 850\,$\mu$m beam). We then re-ran the source extraction method outlined in Section \ref{sec: number_counts} on the map, found how many of our fake sources had been recovered, and hence calculated the completeness for that run. For each run, we included ten fake sources, all with the same peak flux density, and repeated the process 1,000 times for each peak flux density value, starting from 2\,mJy and increasing in steps of 1\,mJy up to 20\,mJy. The final completeness value for each source and peak flux density value is simply the median of the completeness distribution from the 1,000 runs.

\begin{figure}
\centering
\includegraphics[width=0.45\textwidth]{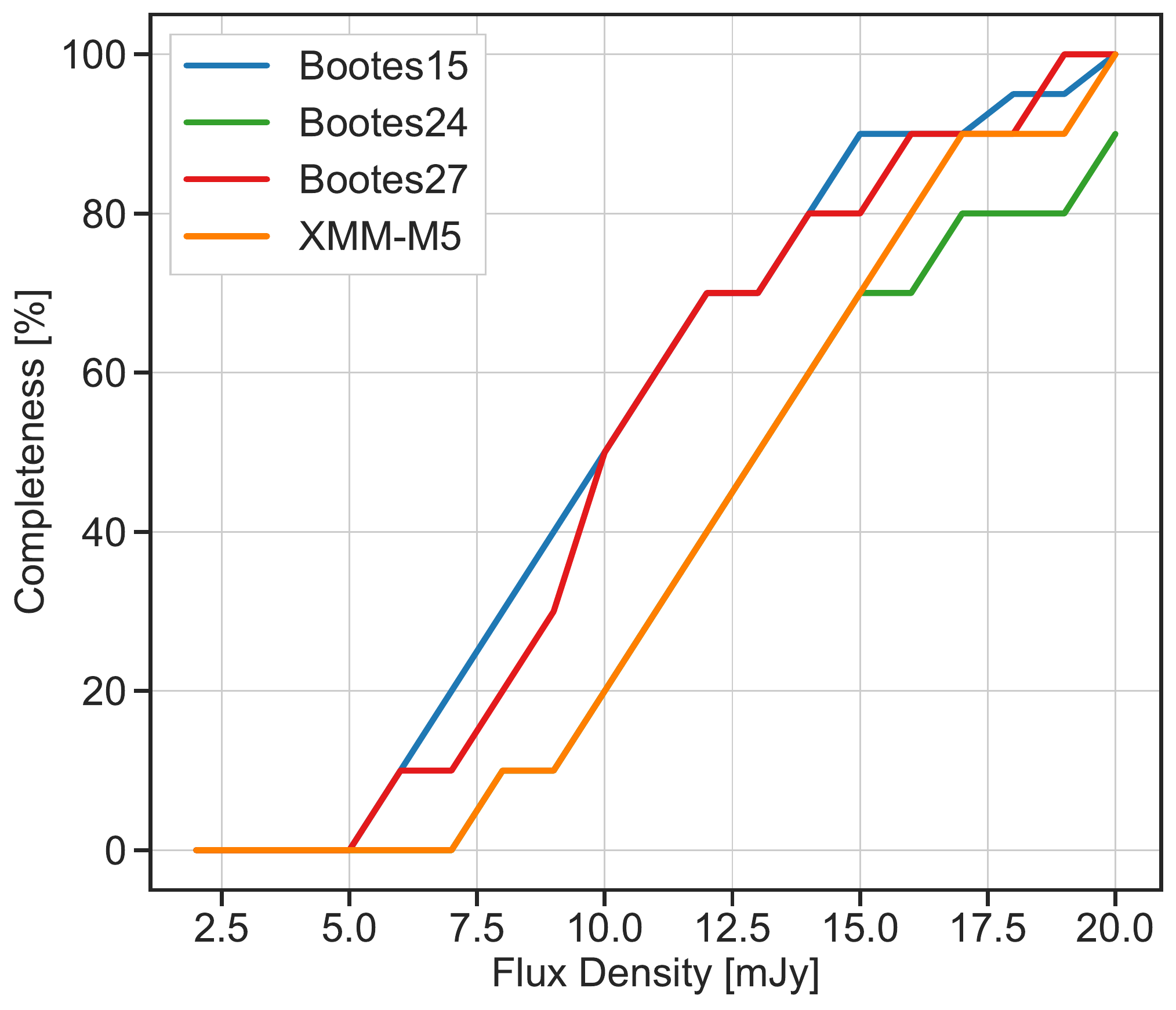}
\caption{The completeness estimates for the SCUBA-2 maps associated with our four 500\,$\mu$m risers. We find that Bootes15 and Bootes27 reach a completeness of $\sim50\%$ at 10\,mJy, and rise to a completeness of $\sim80\%$ at 14\,mJy. Bootes24 and XMM-M5 reach a $\sim80\%$ completeness at 17\,mJy and 16\,mJy respectively.}
\label{fig: completeness_plot}
\end{figure}

In Figure \ref{fig: completeness_plot}, we present the result of this completeness calculation for each of our four SCUBA-2 maps. We find that Bootes15 and Bootes27 both reach a completeness of $\sim 50\%$ at $\sim 10$\,mJy, and rise to a completeness of $\gtrsim 80\%$ at 14\,mJy. Bootes24 and XMM-M5 have a poorer completeness, reaching $\sim 80$\% at $\sim 17$\,mJy and $\sim 16$\,mJy respectively. 

%%%%%%%%%%%%%%%%%%%%%%%%%%%%%%%%%%%%%%%%%%%%%%%%%%

% Don't change these lines
\bsp	% typesetting comment
\label{lastpage}
\end{document}